\documentclass[ALICE,manyauthors]{cernphprep}

\usepackage{lineno}
\usepackage{rotating} 
\usepackage[american]{babel}
\usepackage{verbatim}
\usepackage{makeidx}
\usepackage{amsmath}
\usepackage{slashed}
\usepackage{graphicx}
\usepackage{pstricks}
\usepackage{color, colortbl}
\usepackage{amssymb}
\usepackage{hyperref}
\usepackage{subcaption}	
\usepackage{caption}
\usepackage{bigints}
\usepackage{pdflscape}
\usepackage{setspace}
\usepackage{afterpage}
\usepackage{fancyhdr}
\usepackage{booktabs}
\usepackage{tabu}
\usepackage{multirow}
\usepackage[toc,page]{appendix}
\usepackage{tabularx, booktabs}

\usepackage{color}
\definecolor{White}{rgb}{1,1,1}
\definecolor{LightCyan}{gray}{0.9}
\definecolor{orange}{cmyk}{0.,0.353,1.,0.} 

\begin{document}

\begin{titlepage}
\PHyear{2017}
\PHnumber{192}      
\PHdate{27 July}  
%

\title{Charged-particle multiplicity distributions \\ over a wide pseudorapidity range in proton-proton collisions \\ at $\sqrt{s}=$ 0.9, 7 and 8 TeV}
\ShortTitle{Multiplicity distributions over a wide pseudorapidity range}   

\Collaboration{ALICE Collaboration\thanks{See Appendix~\ref{app:collab} for the list of collaboration members}}
\ShortAuthor{ALICE Collaboration} 

\begin{abstract}
We present the charged-particle multiplicity distributions over a wide pseudorapidity range ($-3.4<\eta<5.0$) for pp collisions at $\sqrt{s}=$ 0.9, 7, and 8 TeV at the LHC.
Results are based on information from the Silicon Pixel Detector and the Forward Multiplicity Detector of ALICE, extending the pseudorapidity coverage of the earlier publications and the high-multiplicity reach.
The measurements are compared to results from the CMS experiment and to PYTHIA, PHOJET and EPOS LHC event generators, as well as IP-Glasma calculations. 
\end{abstract}
\end{titlepage}
\setcounter{page}{2}

\section{Introduction}
The multiplicity of charged particles produced in high-energy pp collisions is one of the key observables to describe the global properties of the interactions and has been the subject of long standing experimental and theoretical investigations.
The pp multiplicity distributions of primary-charged particles have been measured for five increasingly wider pseudorapidity ranges. 
A primary-charged particle is a charged particle with a mean proper lifetime $\tau$ larger than 1 $cm/c$, which is either produced directly in the interaction, or from decays of particles with $\tau$ smaller than 1 $cm/c$, excluding particles produced in interactions with material~\cite{ALICE-PUBLIC-2017-005}.
The results are determined using both the Silicon Pixel Detector (SPD) and the Forward Multiplicity Detector (FMD) in ALICE to widen the pseudorapidity coverage with respect to previous ALICE results \cite{Aamodt:2010ft, Aamodt:2010pp, Adam:2015gka}, which made exclusive use of the SPD.
The extension of the pseudorapidity coverage allows us to increase the high-multiplicity reach of the distributions by around 70-90$\%$ with respect to the previous ALICE publication \cite{Adam:2015gka}, exploring a wider phase space. 

The multiplicity distribution of charged particles produced in high-energy pp collisions is sensitive to the number of interactions between quarks and gluons contained in the protons and to underlying mechanisms of particle production. 
At LHC energies, the particle production is dominated by soft QCD processes, which cannot be treated perturbatively and can only be modeled phenomenologically. 
On the other hand, as the colliding energy grows, the particle production receives increased contributions from hard scattering, which can be treated perturbatively.

We have compared directly our data to previous measurements from CMS \cite{Khachatryan:2010nk}. ATLAS and LHCb use different $p_{\text{T}}$ and $\eta$ ranges \cite{Aad:2010ac,Aaij:2014pza}, making the direct comparison impossible. 
This manuscript presents also an overview of the parameters obtained when fitting multiplicity distributions with the sum of two Negative Binomial Distributions (NBDs).
Additionally, the results have been compared to simulations regularly used at LHC \cite{Skands:2010ak, Bopp:1998rc, Sjostrand:2014zea, Pierog:2013ria} and calculations based on saturation density of gluons in the colliding hadrons \cite{Schenke:2013dpa, McLerran:2015qxa}.

The manuscript is organized in the following way: Sect.~\ref{Sec:detectors} describes the detectors used to measure the charged-particle multiplicity distributions.
Section~\ref{Sec:analysis} explains the analysis procedure in detail.
The systematic uncertainties are described in Sect.~\ref{Sec:systematics} and the results along with comparisons to models are presented in Sect.~\ref{Sec:results}, which contains also the analysis of the NBD fits.
A brief summary and conclusions are finally given in Sect.~\ref{Sec:conc}.

\section{Experimental setup}
\label{Sec:detectors}
Full details of the sub-detectors are given elsewhere  \cite{Aamodt:2008zz}.
ALICE is designed to measure particles over a wide kinematic range $-3.4<\eta<5.0$.
Only the sub-detectors used in this analysis are described, namely the V0 scintillation counters, the SPD, and the FMD.

\subsection{V0 detector}
\label{Sec:v0}
The V0 detector \cite{Abbas:2013taa} is composed of two arrays of 32 scintillators positioned at 330~cm (V0-A) and -90~cm (V0-C) from the nominal interaction point (IP) along the beam axis.
Each array has a ring structure segmented into 4 radial and 8 azimuthal sectors.
The detector has full azimuthal coverage in the pseudorapidity ranges $2.8 < \eta < 5.1$ and $-3.7 < \eta < -1.7$.
The signal amplitudes and times are recorded for each of the 64 scintillators. 
The V0 is appropriate for triggering, thanks to the good timing resolution of each scintillator (1~ns) along with its large acceptance for detecting charged particles.
Different V0 trigger settings are used in the analysis.

\subsection{Silicon Pixel Detector}
\label{Sec:spd}
The SPD are the two innermost cylindrical layers of the ALICE Inner Tracking System (ITS) \cite{Aamodt:2008zz} surrounding the beam line.
The layers have full azimuthal coverage and radii of $3.9$~cm and $7.6$~cm with $9.8\times10^6$ silicon diodes, each of size $50\times425$~${\rm {\mu}m}^2$.
The first layer of the SPD has the largest pseudorapidity coverage of the ITS ($\vert\eta\vert<1.98$ for collisions at the nominal IP).
Besides the readout of individual pixels with signals above a certain threshold, each SPD chip provides a fast signal every 100~ns, indicating a presence of fired pixels (FastOR), making it suitable for triggering.
Charged particles can deposit energy in more than one pixel of the SPD. 
The offline reconstruction combines such neighboring signals into a a single cluster.
The charged-particle multiplicity can then be estimated by counting the number of clusters detected in a SPD layer.
This analysis uses only clusters from the inner layer of the SPD to provide the largest pseudorapidity coverage for particle detection.
Alternatively, clusters from the two SPD layers together with the primary vertex can be combined to form tracklets \cite{Adam:2015gka}, allowing to select primary particles with very high efficiency.
The charged-particle multiplicity is then estimated by counting the number of tracklets.

\subsection{Forward Multiplicity Detector}
\label{Sec:fmd}
The purpose of the FMD is to extend, with high spatial resolution, the charged-particle detection acceptance beyond the reach of the SPD and central detectors in ALICE \cite{Cortese:2004aa}.
The FMD is a silicon strip detector and consists of three sub-detectors placed at 320~cm (FMD1), 79~cm (FMD2), and -69~cm (FMD3) from the nominal IP along the beam pipe.
FMD2 and FMD3 contain both inner and outer rings of silicon strips.
FMD1 is located farther from the IP and has only one inner ring.
Inner rings consist of 10 sensors, each with two azimuthal sectors and 512 strips with radii from 4.2 to 17.2 cm.
Outer rings contain 20 sensors each again with two azimuthal sectors, but with 256 strips, with radii from 15.4 to 28.4 cm.
Each ring (inner or outer), therefore, contains 10240 strips giving in total 51200 strips.
The FMD has full azimuthal coverage in the pseudorapidity ranges $-3.4<\eta<-1.7$ and $1.7<\eta<5.0$.

The FMD records, for each strip, the energy deposited by charged particles traversing the detector.
Various selection criteria, see \cite{Abbas:2013bpa, Adam:2015kda} for details, are applied to the energy measured in each strip to determine if the signal corresponds to a single particle traversing only this strip or also a neighboring strip.
The number of particles traversing the FMD is determined taking into account only the signals which pass these selection criteria.
The majority of the particles which reach the FMD, however, are secondary particles produced in interactions with the beam pipe, the material of the ITS, cables and support structures \cite{Abbas:2013bpa}.
Therefore, a detailed Monte Carlo simulation is needed to determine the number of primary particles produced in the collision.

\section{Analysis procedure}
\label{Sec:analysis}
The multiplicity distribution of the primary particles is affected by many detector effects, such as dead detector regions and secondary particle production.
These detector effects must be minimized and corrected for as they have increasing effects when determining accurately the probability of progressively higher-multiplicity events. 
The unfolding method is used to correct for the detector effects, as will be described in the following. 

\subsection{Event selection}
\label{Sec:evsel}
Collisions at three different center of mass energies (0.9, 7, and 8 TeV) are analyzed.
The data used for the analysis were collected at low beam currents and low pileup during three data taking periods: the 0.9 and 7 TeV samples were acquired in 2010, while the 8 TeV sample was collected in 2012.
The last sample is the most affected by the pileup contamination. For this reason we selected few specific runs with low contamination from pileup events for this energy, and used data taken with interaction rate not exceeding 1 kHz. 
A pileup is defined as more than one collision occurring during the readout time of the detector (300~ns, for the SPD, and 2~$\rm \mu$s sampling time, for the FMD).
Such events produce a bias towards larger multiplicity that enhance mostly the tail of the multiplicity distribution.
Table~\ref{tab:data} shows the number of selected events at each energy and the average number of interactions per bunch crossing, $\langle\mu\rangle$, measured by the experiment \cite{Abelev:2014ffa}.
This parameter is determined experimentally and for this measurements, in which $\langle\mu\rangle \ll 1$, the average probability of having more than one interaction in a single bunch crossing, where at least one interaction occurs, is around $1-2$\% ($\langle\mu\rangle/2$).

\begin{table}[htbp] \centering
\tabulinesep=1.2mm
   \begin{tabu}{l c c} \hline
  $\sqrt{s}$ (TeV)      & Selected MB Events & $\langle\mu\rangle$ \\ \hline\hline
  0.9 					 & $7.4\times10^{6}$               &    $0.04\pm0.01$     \\ \hline
    7       				 & $61\times10^{6}$                &  $0.04\pm0.01$  \\ \hline
  8      					& $26\times10^{6}$ 				&  $0.02\pm0.01$ \\
   \hline
\end{tabu}
\caption{Data samples used in this analysis. For each center-of-mass energy, the total number of selected minimum-bias (MB) events along with the average number of interactions per bunch crossing, $\langle\mu\rangle$, are listed.} \label{tab:data}
\end{table}

Inelastic non-diffractive scatterings are the dominant processes in pp collisions, for which most of the hadrons are produced as a consequence of an exchange of color charge. 
On the contrary, diffractive events can be single-, double-, or central-diffractive.
In Regge theory \cite{Collins:1977jy}, diffraction occurs when the Pomeron interacts with the proton and produces a system of particles, called the diffractive system. 
The case in which only one of the protons dissociates is called single-diffractive.

The signals of the V0 and SPD are used to select events where at least one interaction occurred, which are triggered by requiring the detection of at least one particle in either the V0-A, V0-C, or SPD (MB$_{\text{OR}}$).
Events are divided into three classes depending on further requirements.
The first class includes all inelastic events (the INEL class), which is the same condition as used to select events where an interaction occurred (MB$_{\text{OR}}$ trigger).
The second class (the INEL$>$0) requires the presence of at least one charged particle (tracklet) in the region $\vert\eta\vert<1.0$ in addition to the INEL condition. 
This class has higher trigger efficiency and, therefore, reduced corrections relative to the INEL event class.
The third class requires charged particles to be detected in both the V0-A and the V0-C (MB$_{\text{AND}}$).
This class is used to remove the majority of the single-diffractive events and is, therefore, called the non-single-diffractive (NSD) event class.

To remove interactions of the beam with residual gas in the beam pipe, further selection criteria are applied to the event sample.
Since these interactions can occur anywhere along the beam line, the most efficient way to reject them is to require that the interaction occurs close to the expected bunch-crossing position.
The position of the collision along the beam pipe is determined from the vertex position reconstructed correlating SPD tracklets, with a precision of about 0.2 cm. Beam-gas interactions far from the IP are vetoed by the time difference in the V0-A and V0-C detectors.
The vertex is required to be within 4~cm of the nominal IP position to reduce the contribution from beam-gas interactions and to remove acceptance gaps in the pseudorapidity coverage of the SPD and FMD, since the acceptance depends on the vertex position.

Even though runs with very low $\langle\mu\rangle$ (average 0.04) were chosen, a residual background from pileup events remains.
The majority of pileup events are identified and removed by searching for additional vertices in the same event.
It is required that the uncertainty on the measurement of the longitudinal vertex position is less than 0.2 cm to have the most accurate determination of the vertex. 
Events with an additional vertex separated by more than 0.8~cm from the main one and containing at least 3 attached tracklets are tagged as pileup and removed from the \hbox{analysis}.
Dedicated simulations show that the probability for the pileup event to pass this selection criteria is at most 10$\%$ and the residual pileup does not exceed 10$\%$ up to the highest multiplicities kept in this analysis. 
Therefore, the overall pileup contribution does not exceed 0.2$\%$ for $\langle\mu\rangle=0.04$ and, because it is covered by systematic uncertainties for all multiplicities, no correction is applied for this bias.

\subsection{Unfolding}
\label{Sec:unf}
The FMD had nearly 100\% azimuthal acceptance, but the SPD had a significant number of modules excluded from read-out that must be accounted for.
On the other hand, interactions in detector material increase the detected number of charged particles, in particular in the FMD.
A good understanding of the detector acceptance and of the number of secondary particles which hit the FMD and the SPD is crucial.

\begin{figure}[thbp]
\includegraphics[width=\textwidth]{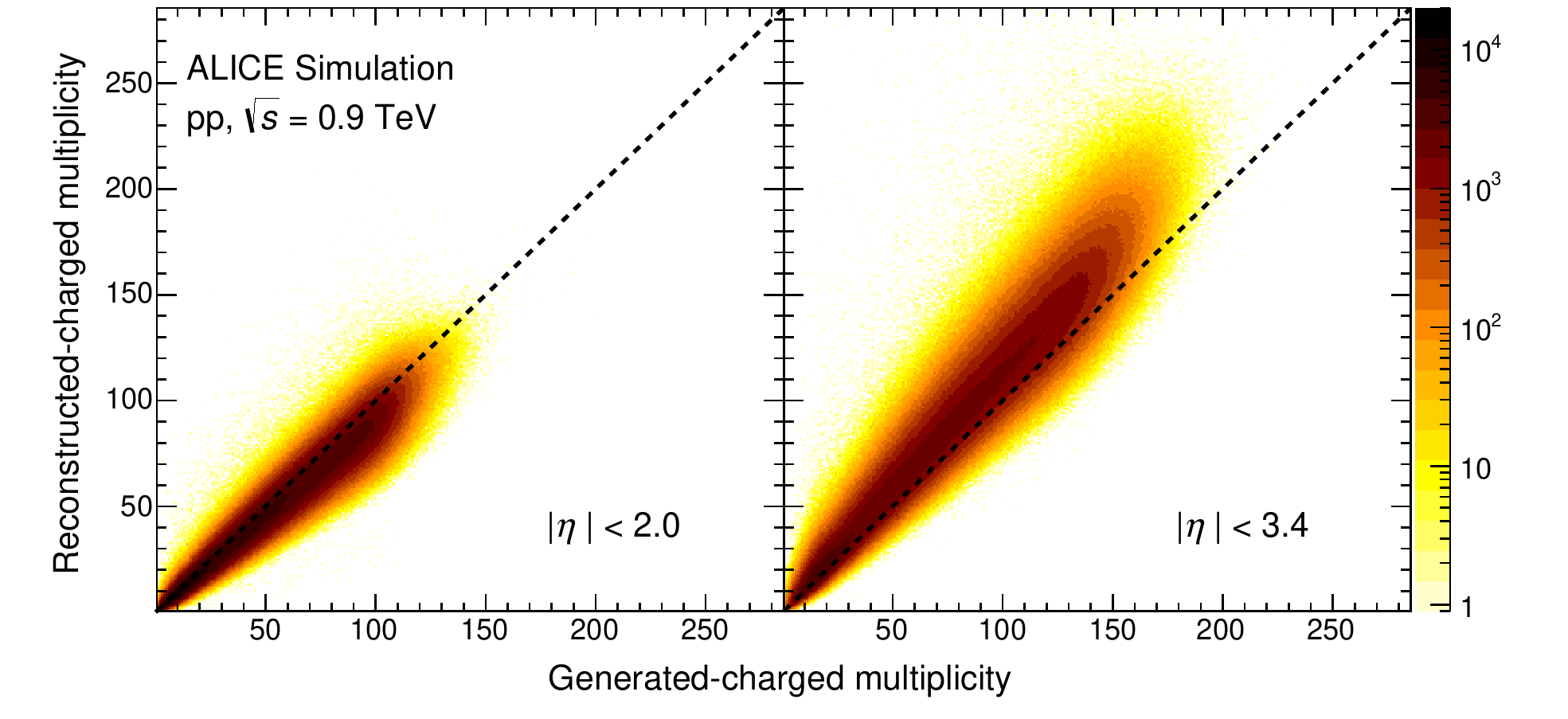}
\caption{Response matrices obtained propagating Monte Carlo generated events, in this case with the PYTHIA ATLAS-CSC flat tune \cite{d'Enterria:2011kw} for the non-single-diffractive event class selection. Left: Matrix including the overlap region between SPD and FMD. Right: Matrix for the region where the majority of the counts are from the FMD. The diagonal (generated=reconstructed) is plotted as a black dotted line.}\label{resp}
\end{figure}

The main ingredients necessary to evaluate the primary multiplicity distributions are the raw (detected) multiplicity distributions and a matrix, which maps the measured multiplicity to the number of charged-primary particles distributions, called true.
The raw multiplicity distributions are determined by counting the number of clusters in the SPD acceptance, the number of signals passing selection criteria in the FMD, or the average between the two if the acceptance of the SPD and FMD overlaps.
The response of the detector is determined by the matrix $R_{mt}$, which corresponds to the probability that an event with true multiplicity $t$ and measured multiplicity $m$ occurs.
This matrix is obtained using PYTHIA ATLAS-CSC flat tune \cite{d'Enterria:2011kw} simulations in which the generated particles are transported through the experimental setup using the GEANT3 \cite{Brun:1994aa} software package. The same reconstruction algorithm is used for simulations of real data.
Experimental conditions and detector settings at the time of data-taking at a center-of-mass energy of $\sqrt{s}=$ 0.9, 7 and 8 TeV are simulated when evaluating the response matrices.
Figure~\ref{resp} shows two different response matrices for different pseudorapidity ranges.
The left panel of Fig.~\ref{resp} shows the response matrix obtained for the $\vert\eta\vert<2.0$.
In this range, the unfolding increases the multiplicity on average because of the acceptance gaps in the SPD.
When the extended pseudorapidity range, $\vert\eta\vert<3.4$, is used, the number of detected counts exceeds on average the number of true counts as the secondary particles in the FMD dominate the bias.
This is shown in the right panel of Fig.~\ref{resp}.

A method based on Bayes' Theorem~\cite{2010arXiv1010.0632D} is used to derive the final multiplicity distributions.
Bayes' Theorem states that the conditional probability $\text{P}(A\vert B)$ (probability of $A$ if $B$ is true) can be written as
\begin{equation}
\text{P}(A\vert B)=\frac{\text{P}(B\vert A)\text{P}(A)}{\text{P}(B)}\,,
\label{bayestheorem}
\end{equation}
in which $\text{P}(A)$ and $\text{P}(B)$ are the independent probabilities of $A$ and $B$, and $\text{P}(B\vert A)$ is the probability of $B$ if $A$ is true. $A$ can be identified as a certain true multiplicity, while $B$ is the measured multiplicity.
The conditional probability $\text{P}(B\vert A)$ is the response matrix of the detector, and can then be computed.

Equation~\ref{bayestheorem} is restated as
\begin{equation}
\widetilde{R}_{tm}=\frac{R_{mt}\text{P}_{t}}{\sum_{\text{t'}}R_{mt'}\text{P}_{t'}}\,,
\end{equation}
where $\text{P}_{t}$ is an a priori guess of the true distribution and $\widetilde{R}_{tm}$ is the matrix of probabilities that allows one to compute the true multiplicity distribution from the measured one.
The unfolded distribution, $U_{t}$, is then obtained from
\begin{equation}
U_{t}=\sum_{m}\widetilde{R}_{tm}M_{m}\,,
\end{equation}
in which $M_{m}$ is the measured distribution.
The obtained $U_{t}$ is used as a priori probability for the next iteration. The number of iterations is fixed to 10. This parameter has been chosen by examining the optimal performance obtained from simulation studies, performing closure tests using different number of iterations.

\subsection{Event selection efficiency}
\label{Sec:eff}
The probability that an event is triggered depends on the multiplicity of charged particles.
At high multiplicities, it is more probable that one of the trigger detectors is fired.
At low multiplicities large trigger inefficiencies for finding events exist and must be corrected for.
The event selection efficiency, $\epsilon_{\text{TRIG}}$, is defined via simulations as
\begin{equation}
\epsilon_{\text{TRIG}}=\frac{N_{\text{ev,reco}}(\text{TRIG}\,\&\,\vert \text{v}_{\text{z,reco}}\vert<4\,{\rm \text{cm}})}{N_{\text{ev,gen}}(\text{TRIG}\,\&\,\vert \text{v}_{\text{{z,gen}}}\vert<4\,{\rm \text{cm}})}\,,
\end{equation}
where the numerator is the number of reconstructed events with the selected hardware trigger condition (MB$_{\text{AND}}$ or MB$_{\text{OR}}$) and with the reconstructed vertex less than 4~cm from the nominal IP, in longitudinal direction. 
There is a dependence in the z vertex distribution and selecting $ \text{v}_{\text{z,reco}}$ introduces a bias in the efficiency.
The effect is visible only for narrow vertex selections, and it is not relevant for $\vert\text{v}_{\text{z}}\vert<4\,{\rm \text{cm}}$.
The denominator is a similar quantity, but for the generated sample (inelastic or non-single-diffractive events).
The unfolded distribution is corrected for the vertex and trigger inefficiency by dividing each multiplicity bin by its $\epsilon_{\text{TRIG}}$ value.

\begin{figure}[htbp]
\begin{subfigure}[h]{0.5\textwidth}
        \includegraphics[width=\textwidth]{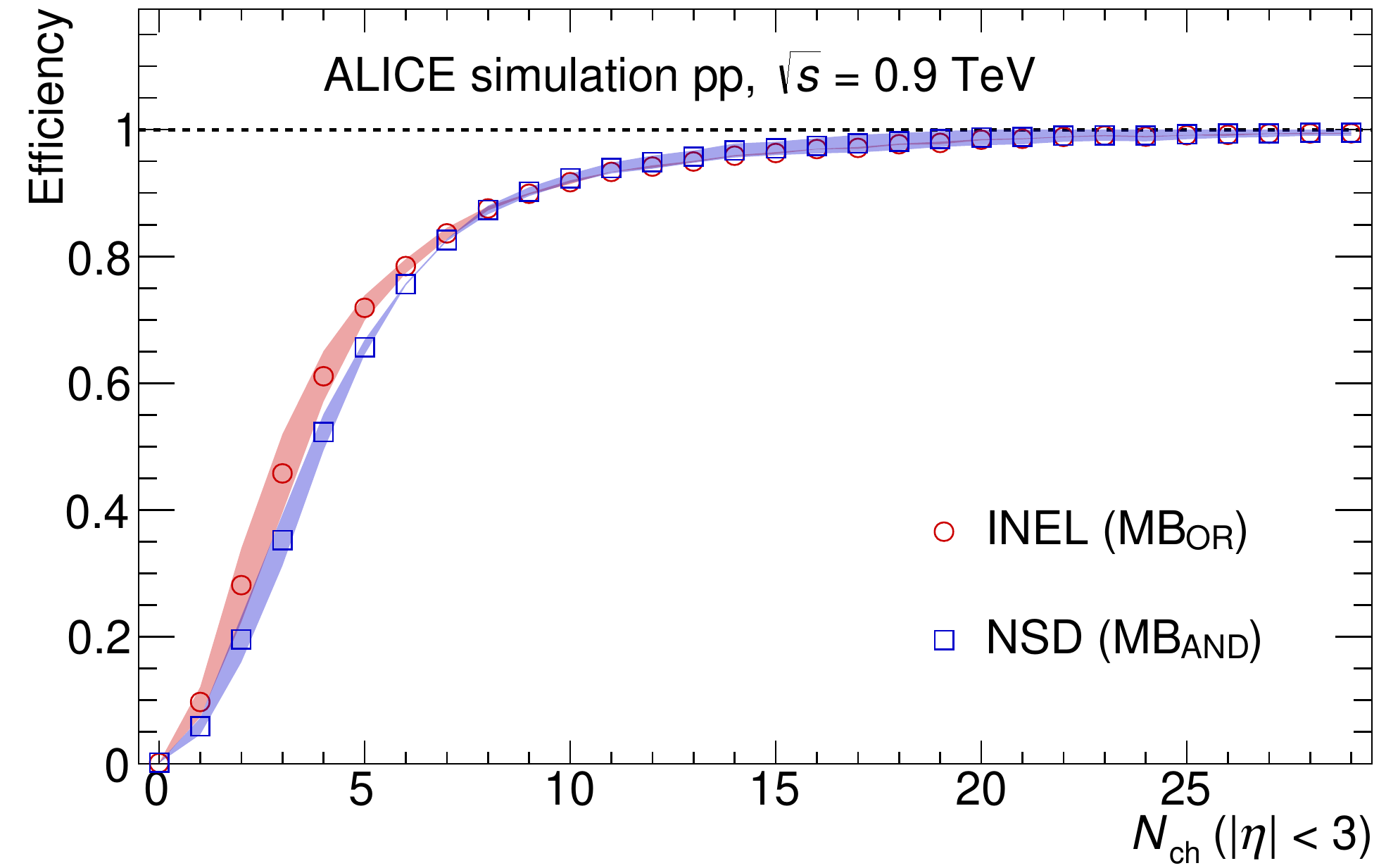}
\end{subfigure}
  \begin{subfigure}[h]{0.5\textwidth}
        \includegraphics[width=\textwidth]{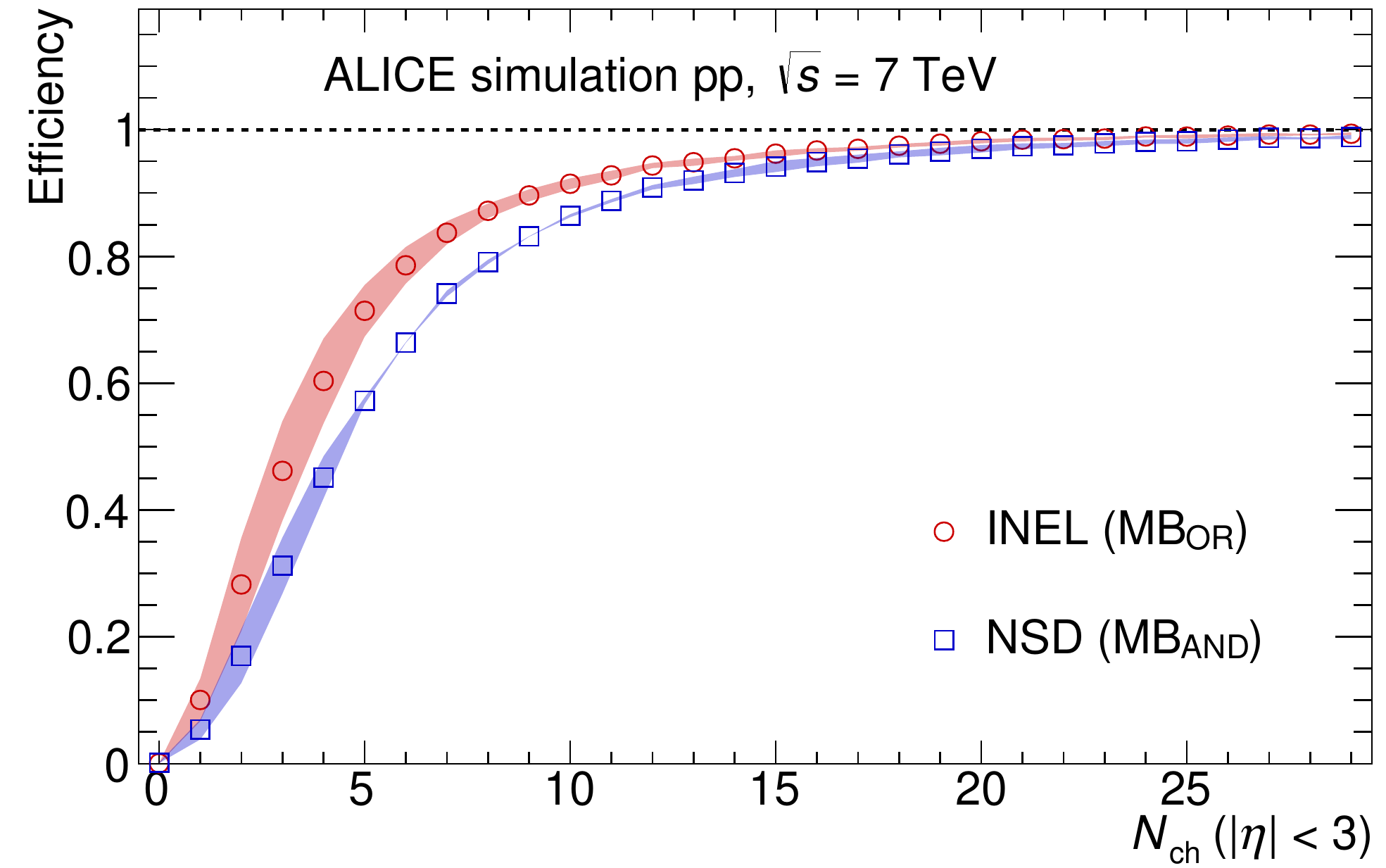}
	\end{subfigure}
\caption{Event selection efficiencies for 0.9 and 7 TeV for both INEL and NSD event samples as a function of the number of primary-charged particles for the $\vert\eta\vert<3.0$ range.}\label{eff}	
\end{figure}

The efficiencies used are shown in Fig.~\ref{eff} for 0.9 and 7 TeV for the range $\vert\eta\vert<3.0$. 
Both the INEL and NSD efficiencies are displayed.
The points are obtained by averaging the efficiencies found with the PYTHIA Perugia 0 \cite{Skands:2010ak} and the PHOJET ~\cite{Bopp:1998rc} diffraction tuned event generators.
Diffraction was accounted for using the Kaidalov-Poghosyan model~\cite{Kaidalov:2009aw} to tune the diffractive processes. 
The event generators are adjusted to reproduce the measured diffraction cross-sections and the shapes of the diffractive masses. The cross-section ratios are $\sigma_{\text{SD}}/\sigma_{\text{INEL}}\backsim0.20$ for upper diffractive mass limit of $M_{\text{X}}<200$ GeV$/c^{2}$, and $\sigma_{\text{DD}}/\sigma_{\text{INEL}}\backsim0.11$ for a pseudorapidity gap of $\Delta \eta>3$, as measured at the LHC~\cite{Abelev:2012sea}. 
The uncertainties are estimated by evaluating the difference between the two event generators and are only relevant at low multiplicity. 
The efficiency of NSD trigger requiring signal in V0-A and V0-C detectors, on both sides of the IP, is lower at low multiplicities than that of INEL trigger, which requires response of at least one V0.
For $N_{\text{ch}}\gtrsim20$ at the widest pseudorapidity ranges probed, both efficiencies reach 100$\%$ and the corresponding systematic uncertainty becomes negligible.

\section{Systematic uncertainties}
\label{Sec:systematics}
The steps involved in the analysis depend on the knowledge of the detector response to charged particles.
The uncertainties in Table~\ref{tab:effsyst} are purely model dependent and related to how diffraction, and soft QCD in general, are processed in the two generators used to determine the efficiency uncertainty. 
The difference between PYTHIA Perugia 0 and PHOJET diffraction tuned generators, used to determine this uncertainty, is larger for small values of $N_{\text{ch}}$.
Therefore, the uncertainty mostly influences the first bins of the multiplicity distributions.
Table~\ref{tab:effsyst} reports the values for charged-multiplicity of 0, 1, and 2.
In general, the Lorentz boost of the diffracted system increases with increasing center-of-mass energies, and single and double diffraction contributions are smaller when going to higher energies.
At wider pseudorapidity ranges there are higher chances of including diffractive events in the distribution.
We observe that the uncertainty for NSD events at $-3.4<\eta<+5.0$ is higher for lower energy in one multiplicity bin, where the description of diffraction differs the most among PYTHIA and PHOJET. 
\begin{table}[htbp] \centering
\tabulinesep=1.1mm
\tabcolsep=3pt
   \begin{tabu}{c l |  c c c | c c c | c c c} \hline
      \multicolumn{11}{c}{Efficiency uncertainty}  \\ \hline\hline
         &     &      &&&&                $\sqrt{s}$ (TeV) &&&        \\ 
        	&     &       & 0.9  &   &    & 7                                                  &  & &   8  \\ \hline
            				&&&&  	 	 & 		& 				  $N_{\text{ch}}$   &&& \\
$\eta$ Range 	& Event Class & 0  & 1  &  2  &  0  &  1   &  2  &  0   &  1  &  2\\    \hline
 			 								& INEL & 15.3 & 6.2 & 3.0 & 27.0 & 11.5 & 5.0 & 28.0 & 12.7 & 8.7 \\
 			 $\vert\eta\vert<2.0$ & NSD & 5.3 & 2.7 & 0.5 & 9.8 & 6.9 & 0.5 & 9.8 & 8.8 & 7.1 \\ 
 										    & INEL$>0$ & -- & 6.2 & 3.1  &	-- & 11.5 & 5.0 & -- & 12.7 & 8.7 \\ \hline
         	                                & INEL & 19.9 & 14.1 & 8.5 & 26.9 & 20.6 & 12.0  & 28.1 & 22.5 & 16.6   \\ 
         	 $\vert\eta\vert<2.4$ &  NSD & 8.1 & 10.7 & 5.7 & 14.4 & 17.6 & 8.5 & 14.6 & 19.8 & 14.2  \\ 
         	 								& INEL$>0$ & -- & 14.1 & 8.5 & -- & 20.6 & 12.0 & -- & 22.5 & 16.6 \\ \hline
 			                                & INEL & 25.8 & 24.4 & 20.7 & 30.4 & 33.5 & 26.0 & 32.1 & 35.2 & 28.9 \\
			  $\vert\eta\vert<3.0$ & NSD & 15.4 & 22.2 & 18.4 & 19.3 & 29.1 & 25.1 & 18.8 & 31.0 & 28.1 \\ 
			  								& INEL$>0$ & -- & 24.4 & 20.7 & -- & 33.5 & 26.0 & -- & 35.2 & 28.9 \\ \hline
                                             &  INEL & 31.2 & 34.5 & 29.6 & 50.4 & 40.3 & 35.4 & 53.4 & 41.4 & 37.2 \\
              $\vert\eta\vert<3.4$  &  NSD & 17.3 & 35.1 & 27.7 & 22.4 & 39.0 & 32.0 & 21.0 & 40.2 & 33.9 \\ 
              								& INEL$>0$ & -- & 34.4 & 29.6 & -- & 40.3 & 35.4 & -- & 41.4 & 37.2 \\ \hline
                                             &  INEL & 48.3 & 45.1 & 36.7 & 71.1 & 43.1 & 44.3 & 75.4 & 45.7 & 47.7  \\
 			 $-3.4<\eta<+5.0$     &  NSD & 35.3 & 64.9 & 34.8 & 55.5 & 31.4 & 39.8 & 50.9 & 34.9 & 43.5 \\ 
 			 								& INEL$>0$ & -- & 45.1 & 36.7 & -- & 43.1 & 44.3 & -- & 45.7 & 47.7 \\ \hline
\end{tabu}
\caption{Systematic uncertainties (in percent) for the efficiency correction, for the INEL, NSD and INEL$>0$ event classes. Numbers are given for multiplicity 0, 1, and 2.} \label{tab:effsyst}
\end{table}

Systematic effects from different sources related to run conditions could produce biases in the number of detected particles.
To investigate such effects, the fluctuations in the results are examined for all three energies by splitting the data set into two separate samples with similar beam conditions, which are then unfolded with two different response matrices.
The response matrices are calculated from simulations relative to the conditions of the runs that are used to unfold.
The two resulting unfolded distributions are then averaged bin by bin.
For \hbox{$\sqrt{s}=$ 0.9} and 8 TeV, the run-to-run fluctuations are found to be negligible up to the value of $N_{\text{ch}}$ in which statistical uncertainties become large.
For~$\sqrt{s}=$ 7 TeV, however, run-to-run fluctuations of around $10-15\%$ in the low $N_{\text{ch}}$ bins are found.

As discussed in Sec.~\ref{Sec:unf}, an accurate detector description is crucial in determining the number of particles created in interactions with detector material in order to retrieve the primary distribution.
For the SPD, little material (besides the beam pipe) exists between the detector and the interaction point, whereas a significant amount of material is present between the FMD and the interaction point.
An estimate of the amount of material versus $\eta$ was done using special satellite collisions \cite{Adam:2015kda}, which occur away from the nominal IP and thereby reduce the amount of traversed material.
The result was that for $-3.4<\eta<-1.7$ the material was underestimated in the simulations of the experiment by a maximum value of 14\%.
For $1.7<\eta<5.0$, the material was estimated with $\pm7$\% precision.  
It is possible to correct for the measurements of the first moment of the distribution (pseudorapidity density \cite{Adam:2015kda}). On the contrary, higher order effects are non-trivial, making it impossible to directly correct in this case for the amount of material.
Instead, the entire raw distribution was unfolded with two response matrices, one which increases the material by 14\% and the other which decreases the material by 7\%.
The difference between the results using the two different matrices determined the maximum systematic error contribution from the material budget.

The last uncertainty was determined by varying the selection criteria used to determine which signals correspond to single particles in the FMD (described in Sec.~\ref{Sec:fmd}) by 5\%.
This value is the maximum variation of the fit parameters to the energy distributions  \cite{Abbas:2013bpa} within each of the three data taking periods. 

The systematic uncertainties from the three sources described along with the one from the efficiency correction are summed in quadrature (see Tab.~\ref{tab:totsyst}).
The methods to determine the systematic uncertainties from material budget give the largest possible variations.
To convert the variations to a root-mean-square value, they have been divided by $\sqrt{3}$ considering that the variation is flat and was taken from the mean bin-by-bin value as the reference, not the full spread.
Three particular multiplicities are considered when reporting the systematic uncertainties in Tab.~\ref{tab:totsyst}: the 0-bin, the mean value $\langle m\rangle$, and the value in which the probability is very low ($\text{P}(N_{\text{ch}})=10^{-4}$).
For $\text{P}(N_{\text{ch}})<10^{-4}$ the uncertainties grow rapidly and the level of the systematic uncertainty depends strongly on the multiplicity.
The $\text{P}(N_{\text{ch}})$ region with lowest uncertainty is near the mean of the distribution.
\begin{table}[htbp] \centering
\tabulinesep=1.1mm
\tabcolsep=3pt
\small
   \begin{tabu}{ c l | c c c | c c c | c c c} \hline
         \multicolumn{11}{c}{Total uncertainty }  \\ \hline\hline
          			&   		&    &   & & & $\sqrt{s}$ (TeV) & & & &   \\ 
       	&     & & 0.9  & & & 7 & & &   8  \\ \hline
            				  	 	 & 		& 				&  & &$N_{\text{ch}}$  & $N_{\text{ch}}$ & $\text{P}(N_{\text{ch}})$& & &   \\
            	$\eta$ Range 						& Event Class & \footnotesize{0}  &  \footnotesize{$\langle m\rangle$}  &  \footnotesize{$10^{-4}$}  &  \footnotesize{0}  &  \footnotesize{$\langle m\rangle$}  &  \footnotesize{$10^{-4}$} &    \footnotesize{0}  &  \footnotesize{$\langle m\rangle$}  &  \footnotesize{$10^{-4}$} \\    \hline
			 							  & INEL & 16.6 & 2.0 & 12.2 & 27.4 & 2.5 & 28.8 & 27.1 & 2.9 & 17.6 \\
 			 $\vert\eta\vert<2.0$ & NSD & 7.6 & 1.3 & 12.3 & 11.0 & 4.4 & 30.2 & 9.8 & 4.6 & 19.0  \\ 
 											& INEL$>$0 &  --  & 1.9 & 12.9 & -- & 4.0 & 30.2 & -- & 4.2 & 17.9  \\ \hline
         	 	 							  & INEL  & 20.8 & 2.3 & 21.2 & 27.3 & 2.9 & 37.7 & 28.1 & 2.2 & 34.6   \\ 
         	 $\vert\eta\vert<2.4$ &  NSD & 10.0 & 3.1 & 18.0 & 15.1 & 4.4 & 39.2 & 14.6 & 7.4 & 36.9  \\ 
         	 								& INEL$>$0 &  --  & 2.7 & 22.2 & -- & 4.6 & 40.1 & -- & 7.4 & 35.8 \\ \hline
 			   							 & INEL & 26.2 & 1.8 & 26.1 & 30.8 & 4.9 & 43.9 & 32.1 & 4.8 & 43.6  \\
			  $\vert\eta\vert<3.0$ & NSD & 16.1 & 2.6 & 26.8 & 19.9 & 5.6 & 44.2 & 18.8 & 7.0 & 43.8  \\ 
			 								& INEL$>$0 &  --  & 2.7 & 26.9 & -- & 5.9 & 52.8 & -- & 6.7 & 45.0 \\ \hline
              								  &  INEL & 31.5 & 2.2 & 27.3 & 50.8 & 6.1 & 51.9 & 53.4 & 5.3 & 50.1 \\
              $\vert\eta\vert<3.4$ &  NSD & 17.8 & 2.4 & 28.6 & 23.1 & 6.1 & 53.5 & 21.1 & 6.1 & 52.9 \\ 
              								& INEL$>$0 &  --  & 2.4 & 28.7 & -- & 5.9 & 52.8 & -- & 5.9 & 51.5 \\ \hline
             							    &  INEL & 48.4 & 1.3 & 29.8 & 71.5 & 5.4 & 60.5 & 75.4 & 3.9 & 57.4  \\
 			  $-3.4<\eta<+5.0$  &  NSD & 35.6 & 2.1 & 30.7 & 56.0 & 6.9 & 63.8 & 50.9 & 7.0 & 61.5 \\ 
 			 								& INEL$>$0 &  --  & 2.4 & 30.8 & -- & 5.7 & 63.4 & -- & 5.7 & 59.9 \\ \hline           
\end{tabu}
\caption{Total systematic uncertainties (in percent), for the INEL, NSD  and INEL$>$ 0 event classes. Numbers are given at multiplicity values of 0, the mean $\langle m\rangle$, and when $\text{P}(N_{\text{ch}})=10^{-4}$.} \label{tab:totsyst}
\end{table}

\section{Results}
\label{Sec:results}
The multiplicity distributions have been measured for the three event classes (INEL$>$0, INEL, and NSD) for pp collisions at $\sqrt{s}=$ 0.9, 7, and 8 TeV.
To extract the relative contributions from hard and soft processes, the distributions are fitted with double Negative Binomial Distributions.
The results are also compared with the LHC measurements done by CMS and with distributions obtained from models including the IP-Glasma, which is based on the Color Glass Condensate (CGC)~\cite{Iancu:2003xm}.

\begin{figure}[htbp]
    \begin{subfigure}[c]{0.5\textwidth}
        \centering
        \includegraphics[width=\textwidth]{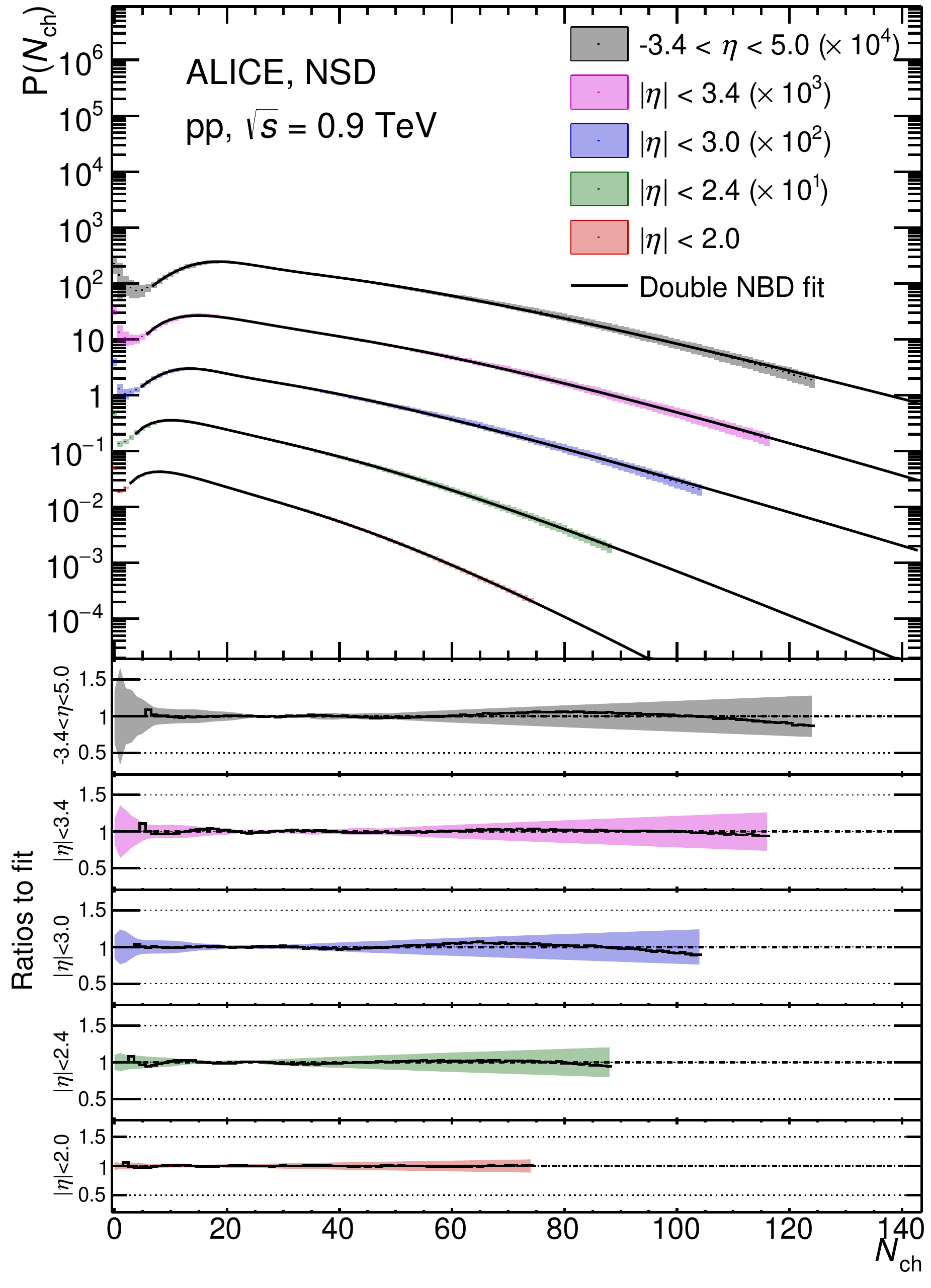}
    \end{subfigure}
    \begin{subfigure}[c]{0.5\textwidth}
        \centering
        \includegraphics[width=\textwidth]{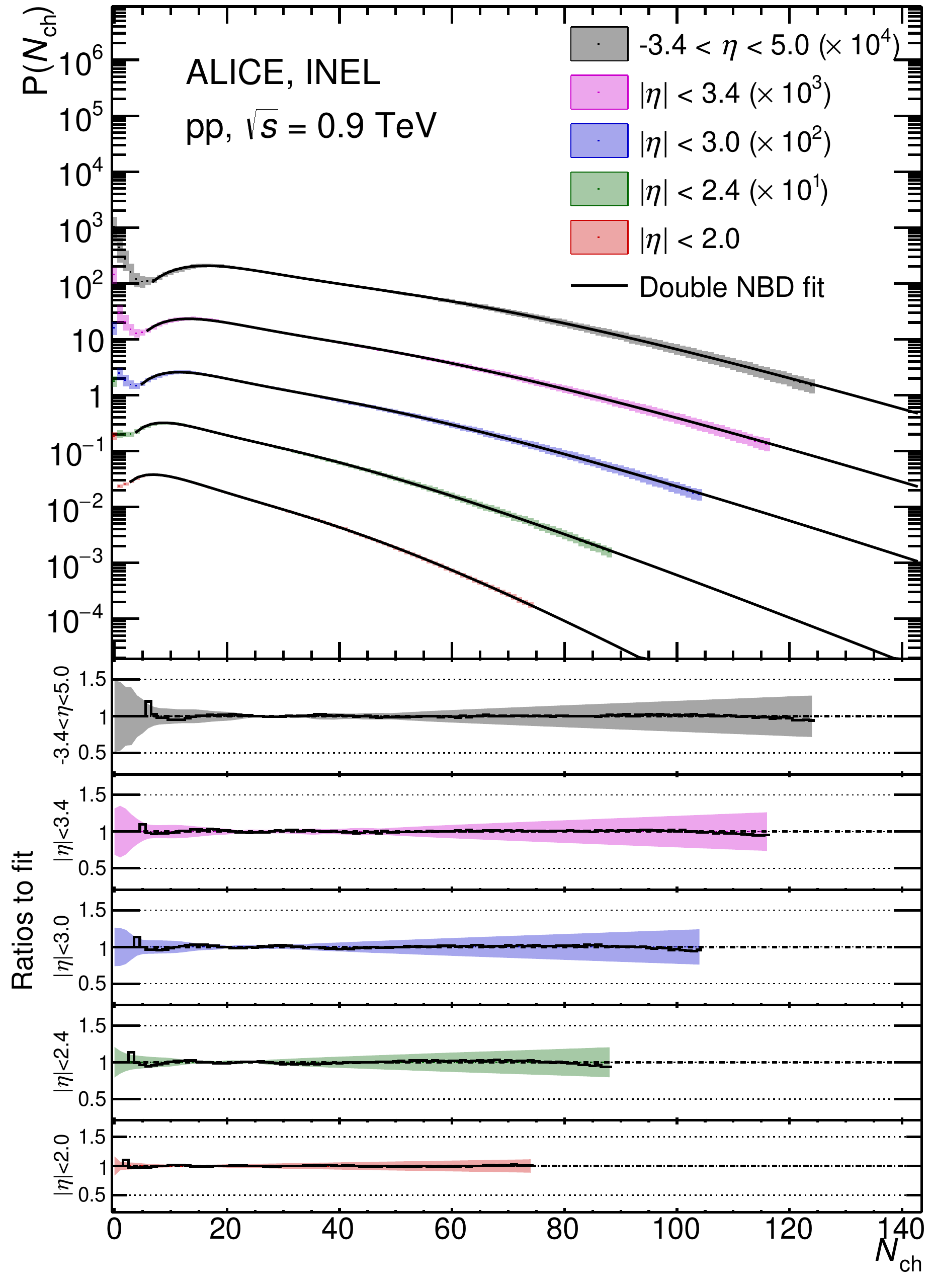}
	\end{subfigure} 
	    \begin{subfigure}[c]{0.5\textwidth}
        \centering
        \includegraphics[width=\textwidth]{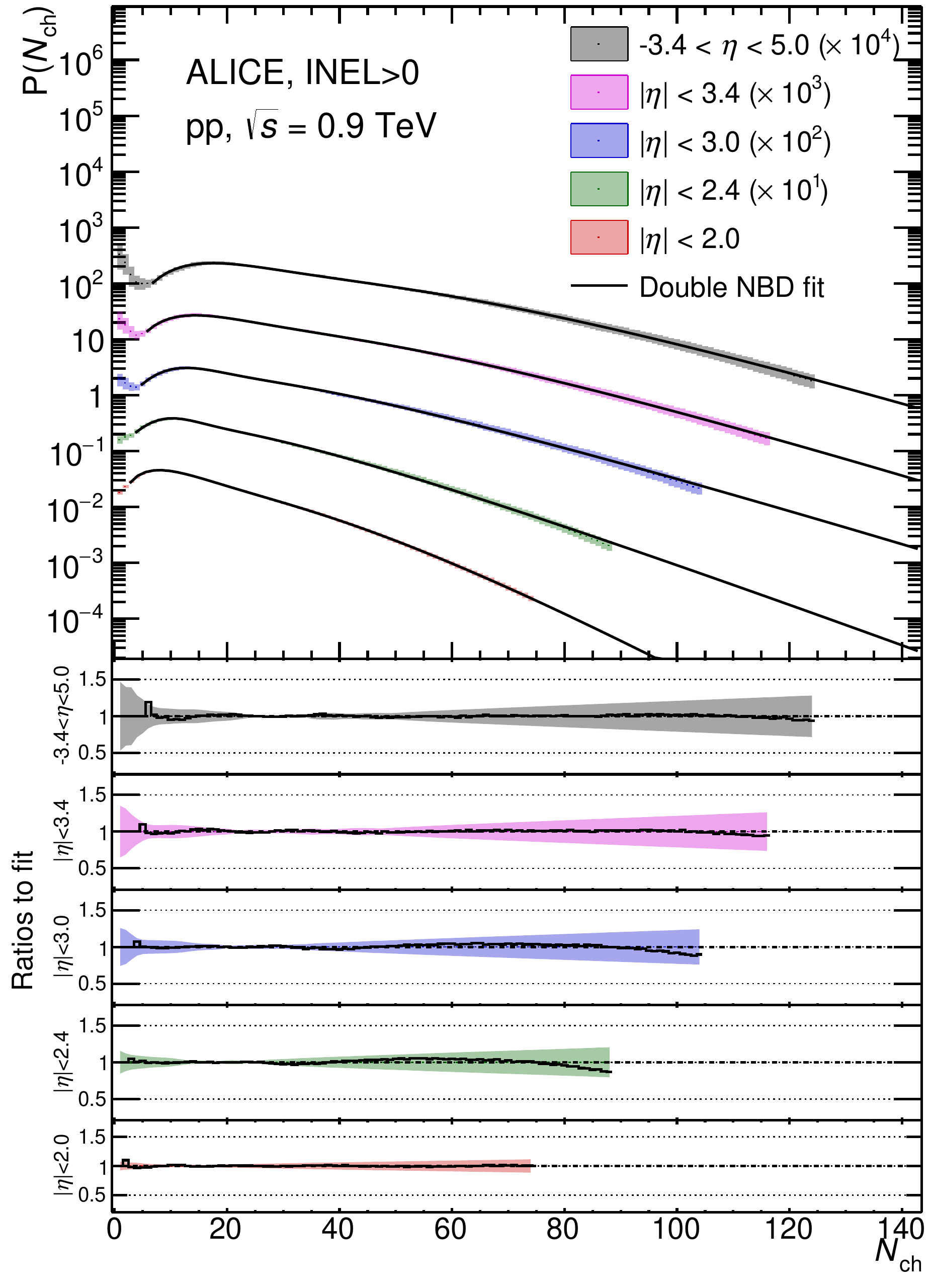}
	\end{subfigure} 
\caption{Charged-particle multiplicity distributions for NSD (top left), INEL (top right) and INEL$>$0 (bottom) pp collisions at $\sqrt{s}=0.9$ TeV.
The lines show fits to double NBDs.
Ratios of the data to the fits are also presented.
Combined systematic and statistical uncertainties are shown as bands.}
\label{V0AND900}
\end{figure}

\begin{figure}[htbp]
    \begin{subfigure}[c]{0.5\textwidth}
        \centering
        \includegraphics[width=\textwidth]{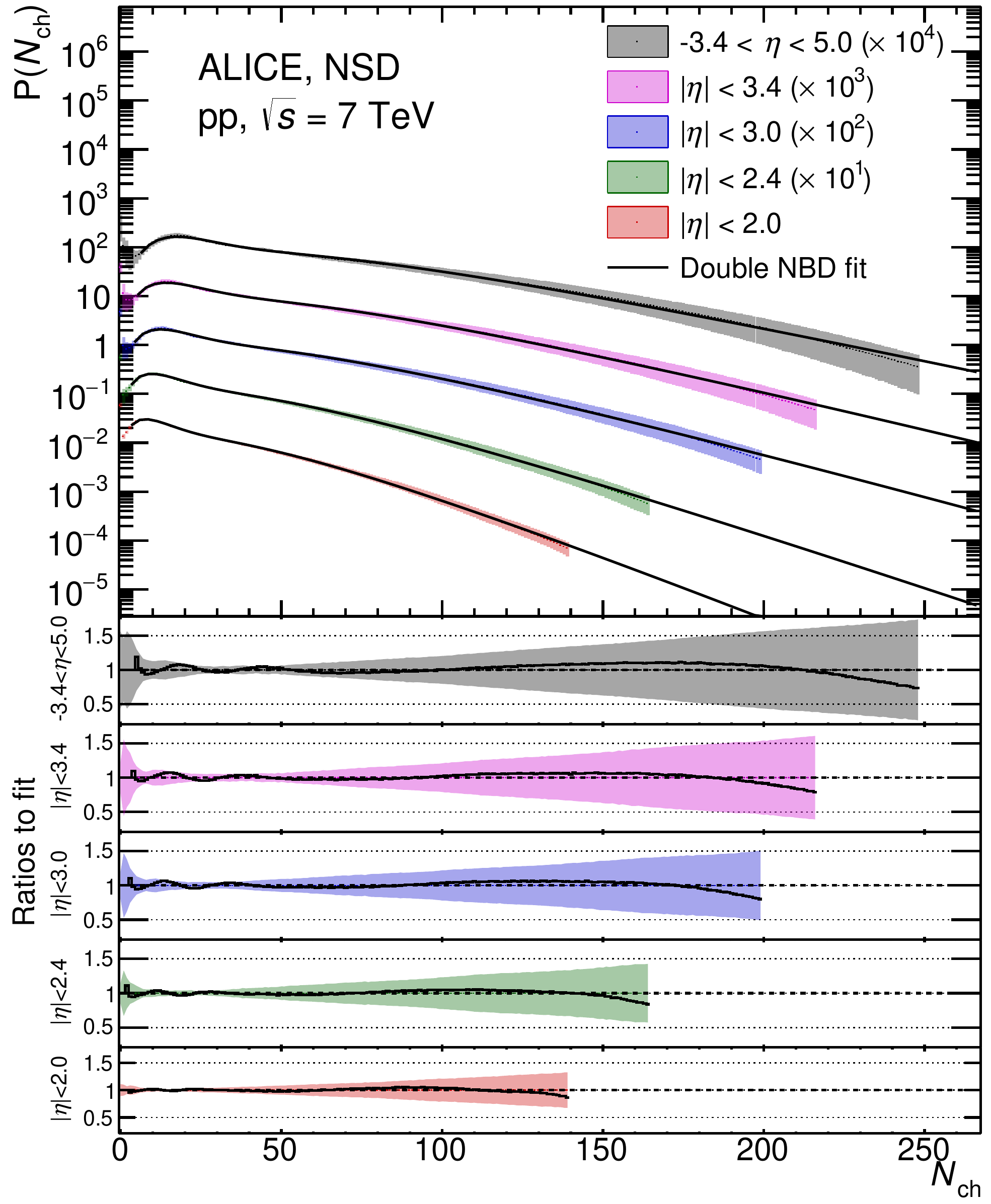}
    \end{subfigure}
    \begin{subfigure}[c]{0.5\textwidth}
        \centering
        \includegraphics[width=\textwidth]{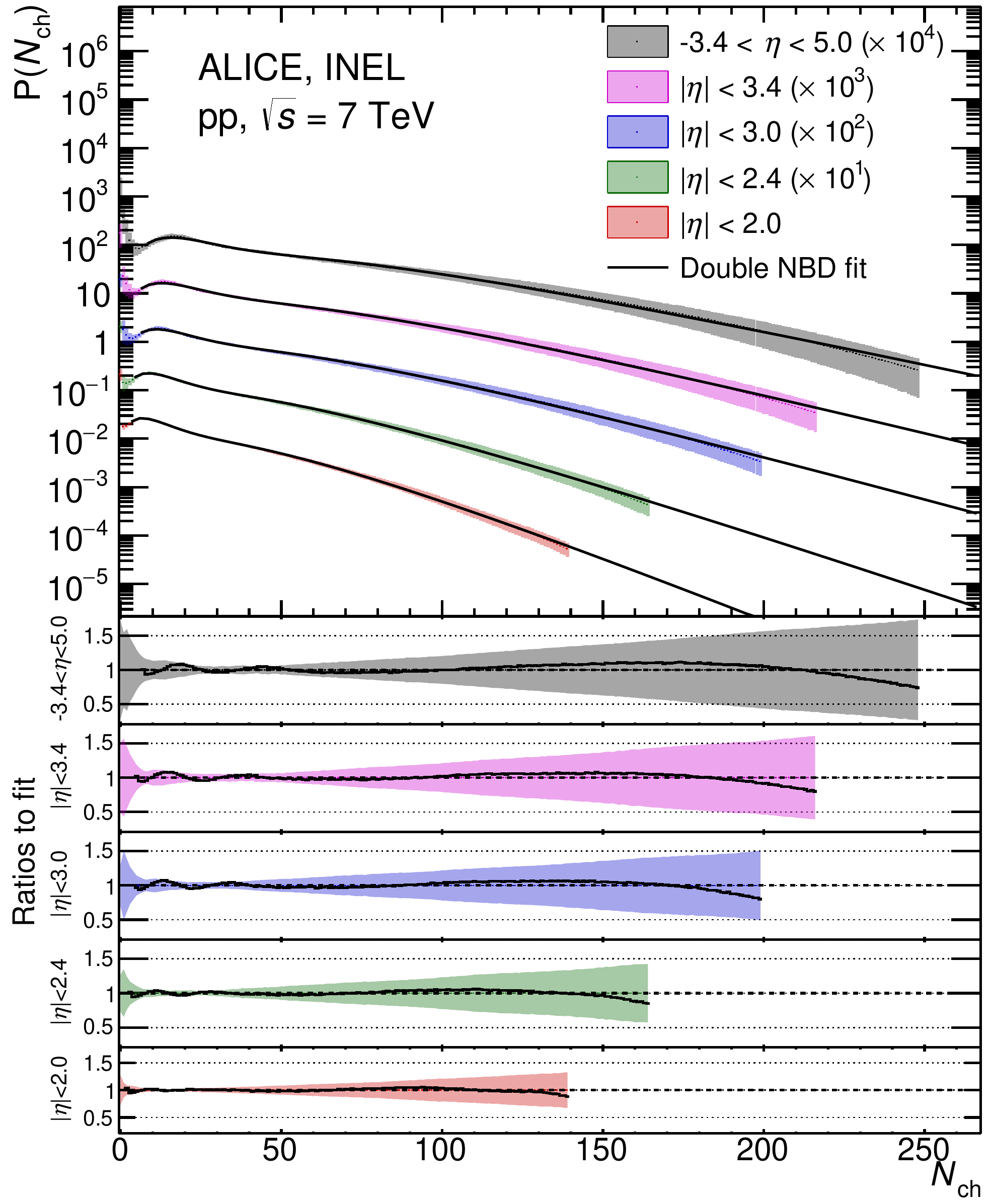}
	\end{subfigure} 
	    \begin{subfigure}[c]{0.5\textwidth}
        \centering
        \includegraphics[width=\textwidth]{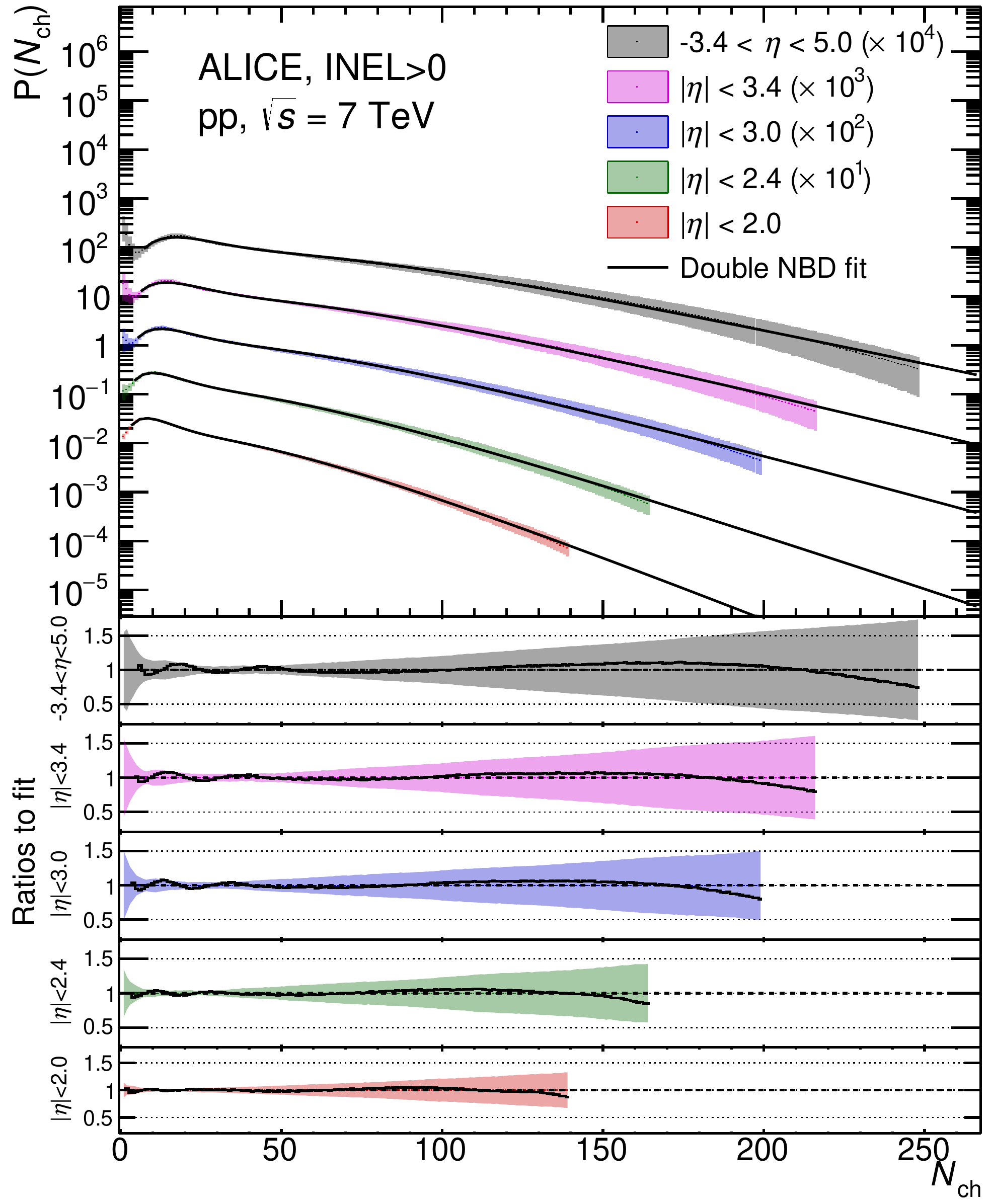}
	\end{subfigure} 
\caption{Charged-particle multiplicity distributions for NSD (top left), INEL (top right) and INEL$>$0 (bottom) pp collisions at $\sqrt{s}=7$ TeV.
The lines show fits to double NBDs.
Ratios of the data to the fits are also presented.
Combined systematic and statistical uncertainties are shown as bands.}
\label{V0AND7000}
\end{figure}

\begin{figure}[htbp]
    \begin{subfigure}[c]{0.5\textwidth}
        \centering
        \includegraphics[width=\textwidth]{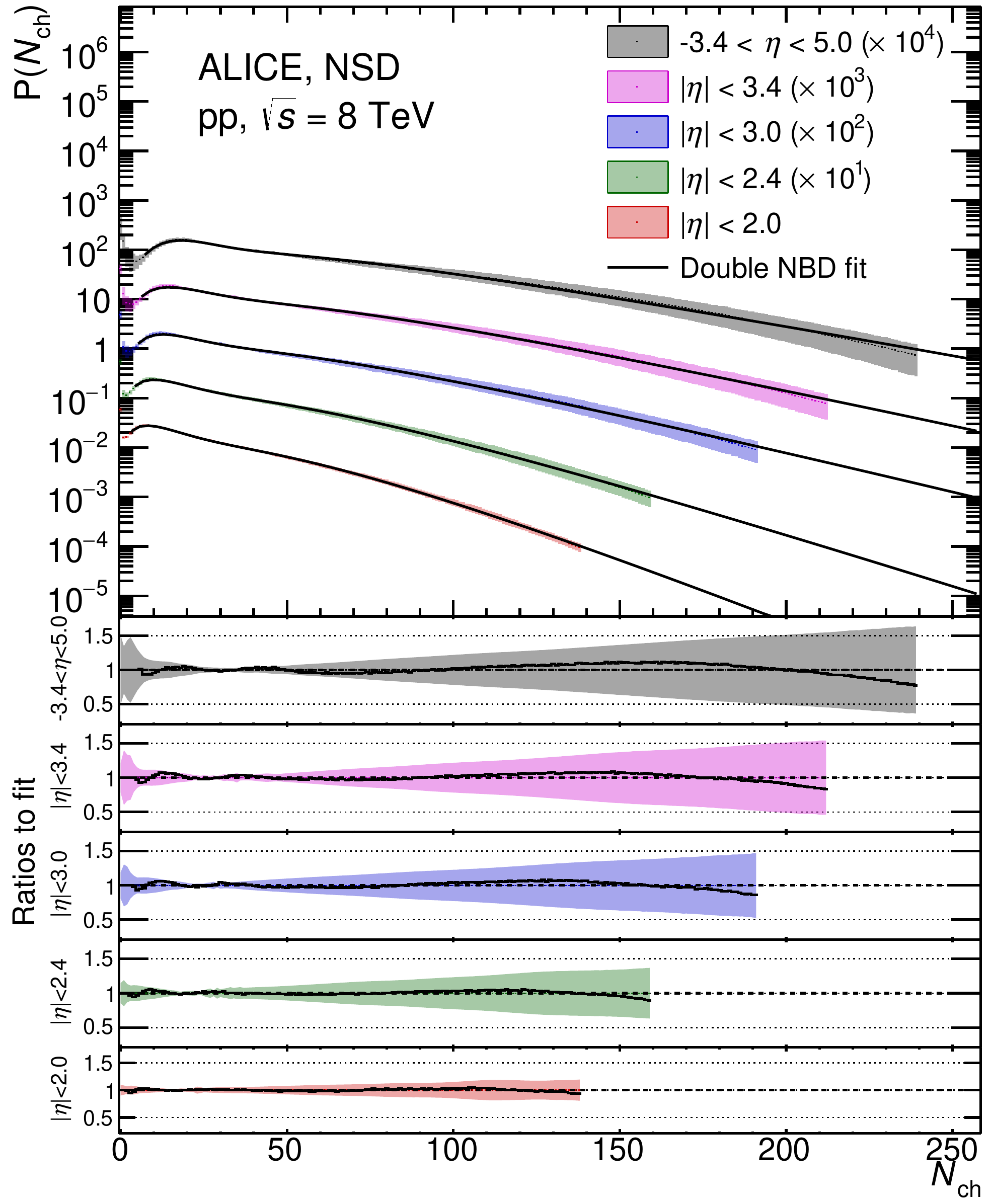}
    \end{subfigure}
    \begin{subfigure}[c]{0.5\textwidth}
        \centering
        \includegraphics[width=\textwidth]{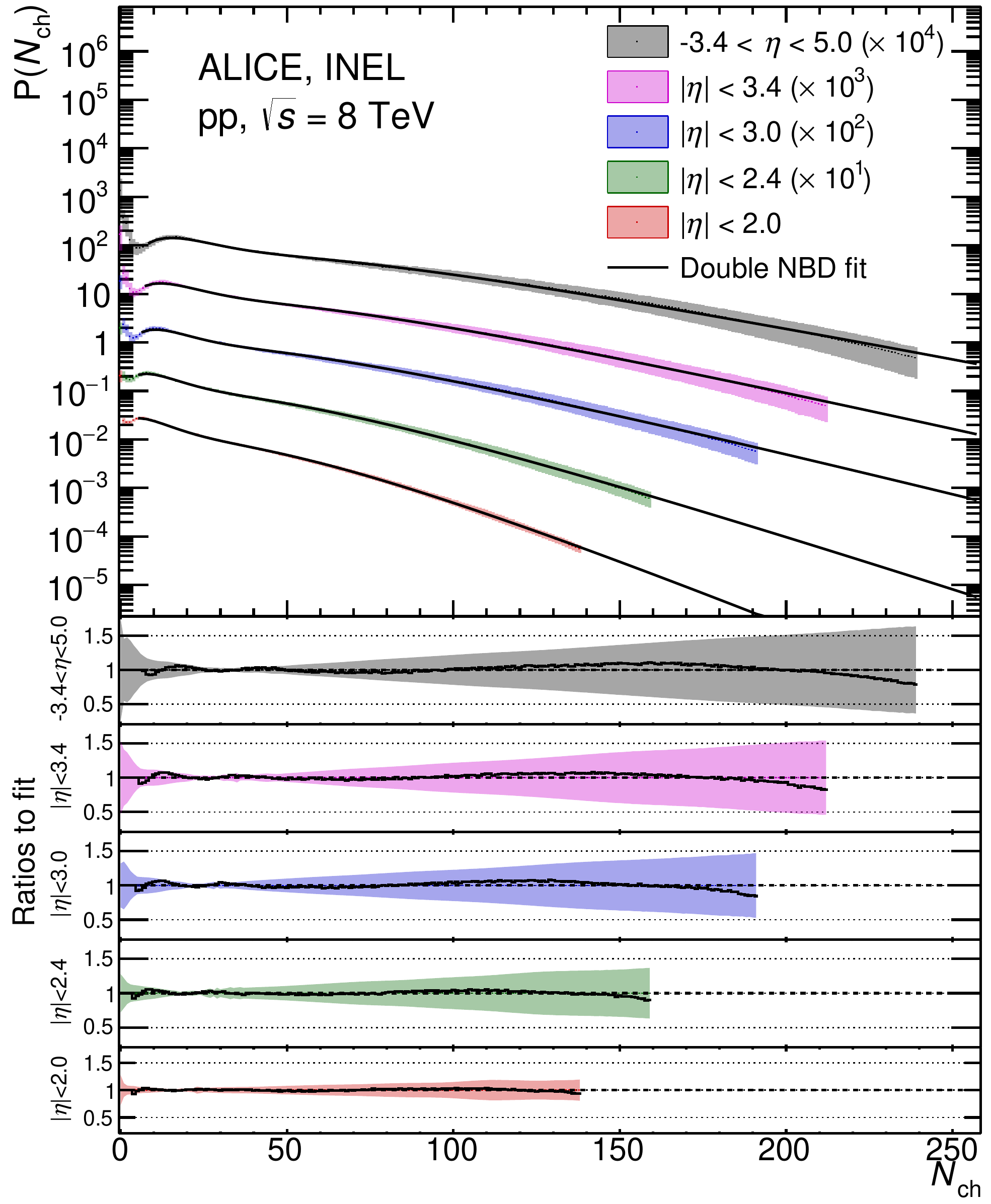}
	\end{subfigure} 
	    \begin{subfigure}[c]{0.5\textwidth}
        \centering
        \includegraphics[width=\textwidth]{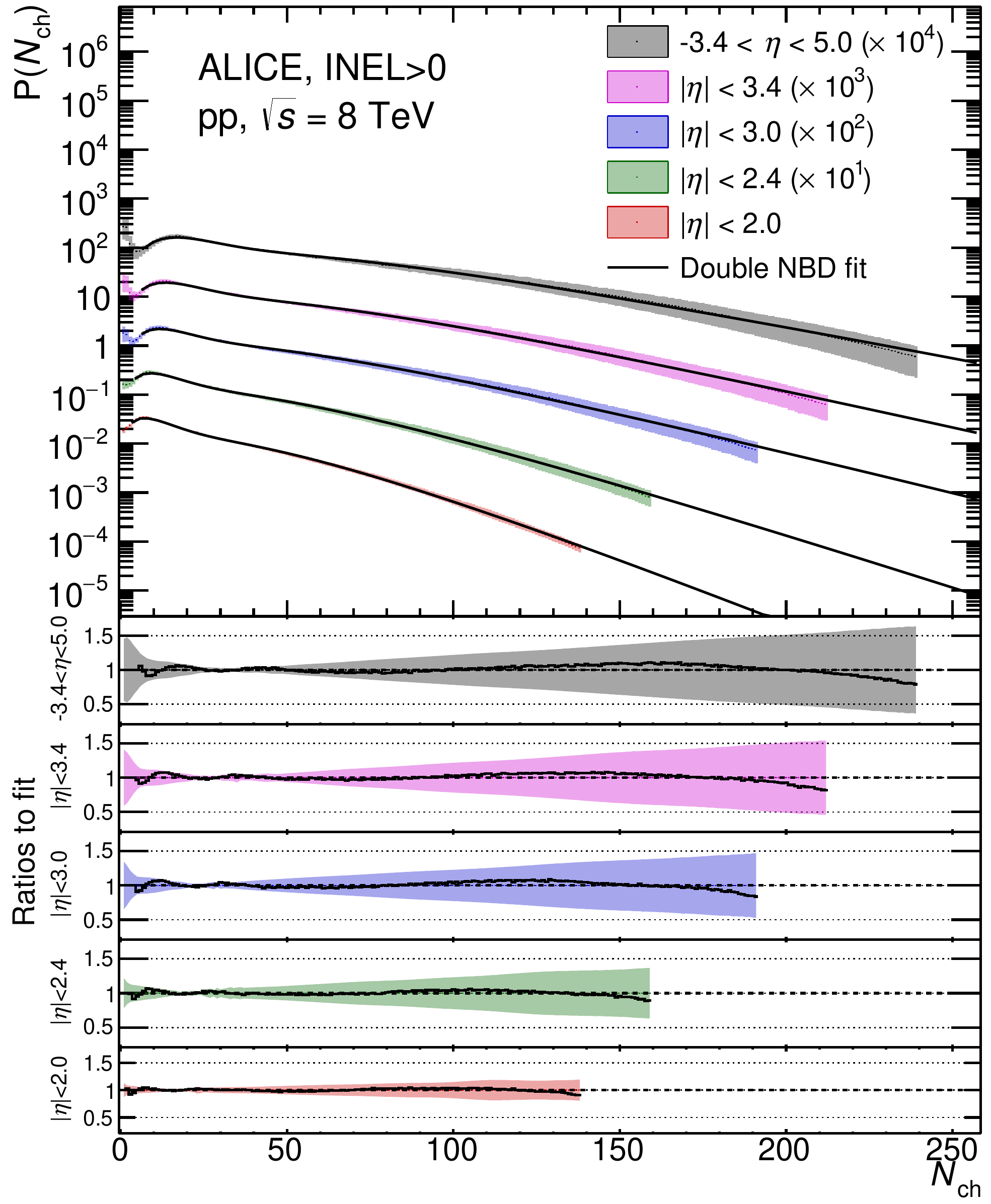}
	\end{subfigure} 
\caption{Charged-particle multiplicity distributions for NSD (top left), INEL (top right) and INEL$>$0 (bottom) pp collisions at $\sqrt{s}=8$ TeV.
The lines show fits to double NBDs.
Ratios of the data to the fits are also presented.
Combined systematic and statistical uncertainties are shown as bands.}
\label{V0AND8000}
\end{figure}

\subsection{Multiplicity distributions}
\label{Sec:multdists}
In Figs.~\ref{V0AND900}, ~\ref{V0AND7000} and ~\ref{V0AND8000}, the obtained multiplicity distributions for 0.9, 7 and 8 TeV with NSD, INEL and INEL$>$0 triggers are shown for five pseudorapidity ranges, $\vert\eta\vert<2.0$, $\vert\eta\vert<2.4$, $\vert\eta\vert<3.0$, $\vert\eta\vert<3.4$ and $-3.4<\eta<+5.0$. The colored bands represent the combined systematic and statistical uncertainties. 
The distributions are scaled by powers of 10 for clarity.
The lines show fits to double NBDs, explained in Sect.~\ref{Sec:nbd}.
Extending the pseudorapidity coverage with respect to the previous ALICE publications \cite{Aamodt:2010ft,Aamodt:2010pp,Adam:2015gka} allows to extend the high-multiplicity reach.

\begin{figure}[htb]
    \begin{subfigure}[c]{0.49\textwidth}
        \centering
        \includegraphics[width=\textwidth]{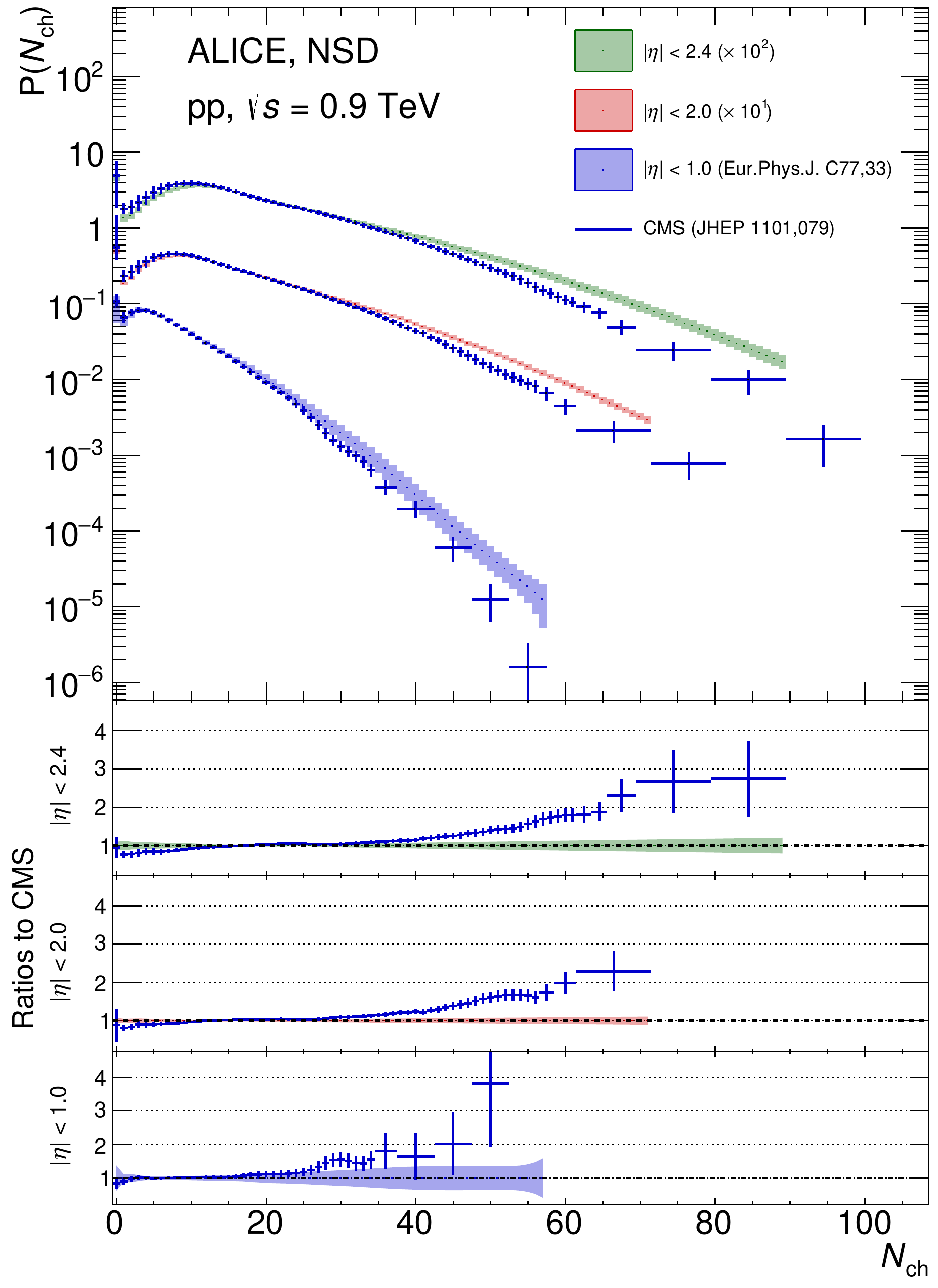}
    \end{subfigure}
    \begin{subfigure}[c]{0.51\textwidth}
        \centering
        \includegraphics[width=\textwidth]{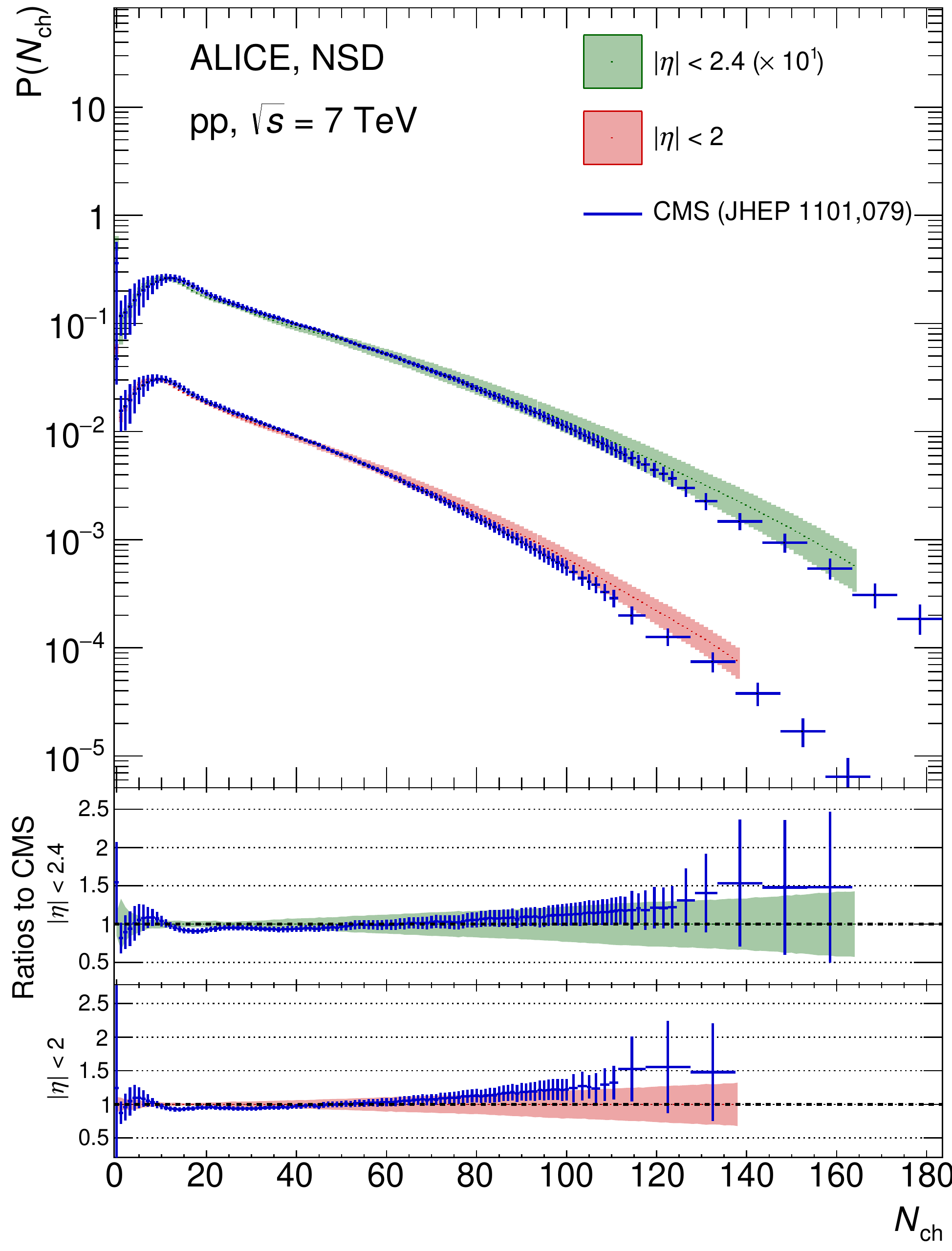}
	\end{subfigure} 
\caption{Comparison of the multiplicity distributions for NSD pp collisions at $\sqrt{s}=0.9$ TeV (left) and 7 TeV (right) with CMS~\cite{Khachatryan:2010nk} measurements in the same pseudorapidity ranges and previous ALICE measurements~\cite{Adam:2015gka}. Combined systematic and statistical uncertainties are shown as bands.}
\label{ALICECMSMineV0AND900}
\end{figure}

In Fig.~\ref{ALICECMSMineV0AND900} (left panel), the comparison to published CMS data~\cite{Khachatryan:2010nk} is shown for the NSD event class at $\sqrt{s}=0.9$ TeV in three $\eta$ ranges.
Discrepancies are observed in the comparison with CMS, both in the first bins and, especially, in the tails.
Differences in the first bins can be attributed presumably to the different models used to describe the diffraction masses.
In particular, CMS is not using any diffraction tuned simulation.
Moreover, in this paper, single-diffractive events include a cut on diffractive mass~\cite{Kaidalov:2009aw}, which is different from what CMS has used.
In the tails, the CMS data are systematically lower, due to normalization. 
This behavior was also observed when comparing CMS distributions to those obtained in a narrower pseudorapidity range where only SPD tracklets are used \cite{Adam:2015gka} as shown in the distribution for $\vert\eta\vert<1.0$ in Fig.~\ref{ALICECMSMineV0AND900}.

In the right panel of Fig.~\ref{ALICECMSMineV0AND900}, the comparison to CMS at 7 TeV~\cite{Khachatryan:2010nk} is shown.
Good agreement with CMS is observed except in the very first bins presumably due to the different treatment of diffraction masses.
The measurement reported in this manuscript, performed with the SPD clusters, agrees with the analysis performed on SPD tracklets \cite{Adam:2015gka} within \hbox{systematic uncertainties}.

\begin{figure}[htb]
    \begin{subfigure}[c]{0.5\textwidth}
        \centering
        \includegraphics[width=\textwidth]{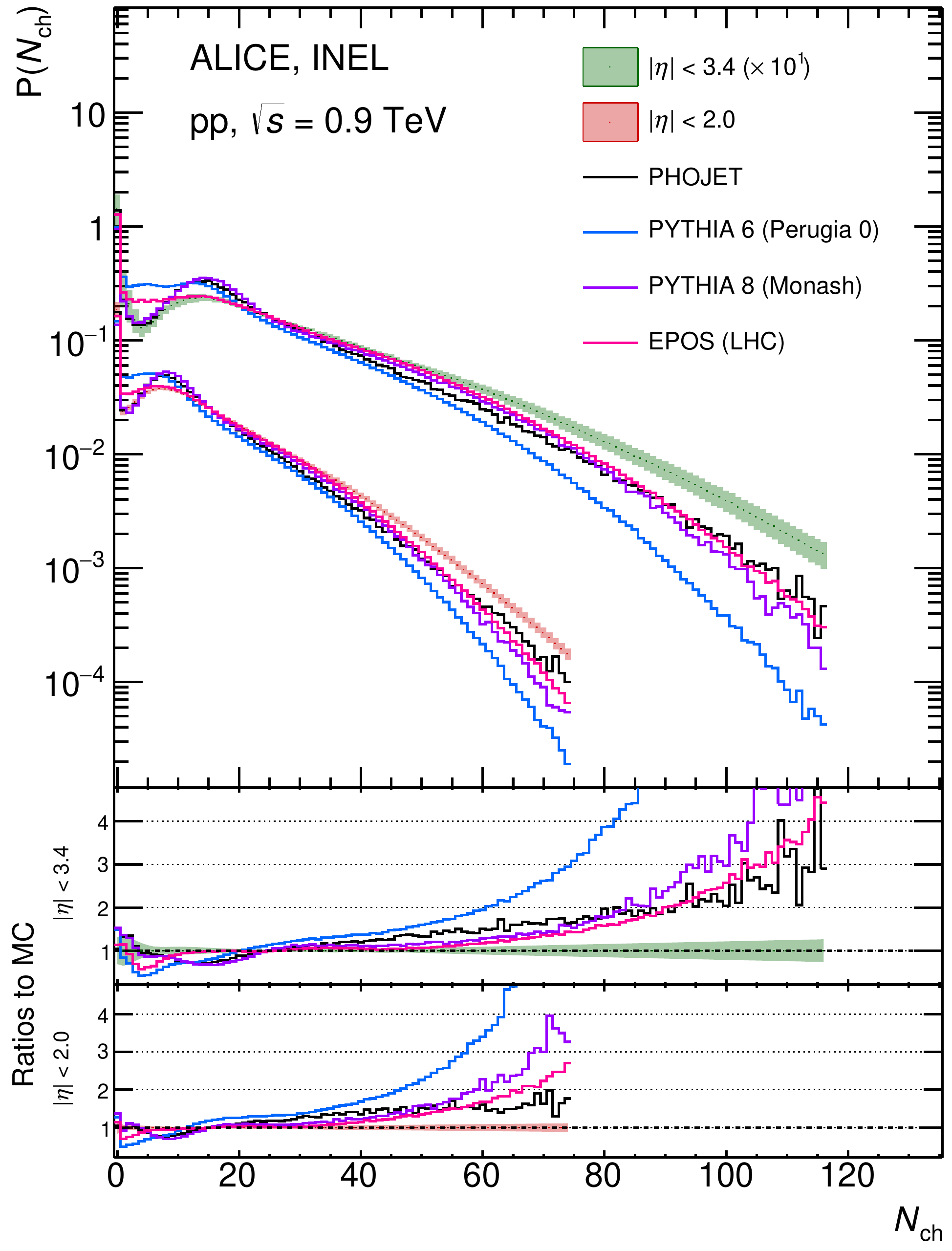}
    \end{subfigure}
    \begin{subfigure}[c]{0.5\textwidth}
        \centering
        \includegraphics[width=\textwidth]{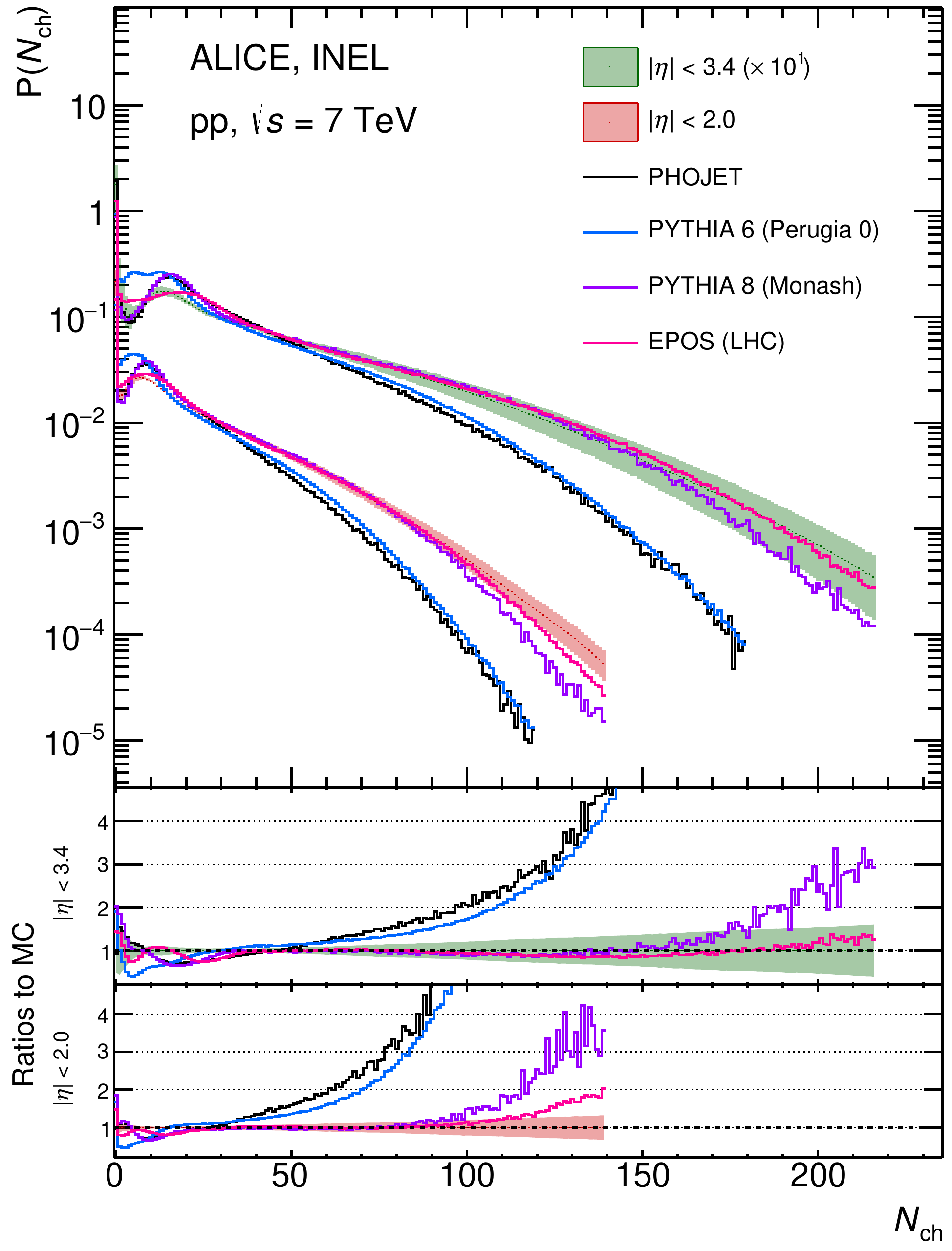}
	\end{subfigure}
\caption{Comparison of multiplicity distributions for INEL events to PYTHIA 6 Perugia 0, PYTHIA 8 Monash, PHOJET and EPOS LHC at 0.9 (left) and 7 TeV (right). Combined systematic and statistical uncertainties are shown as bands.}
\label{MCMineINEL900}
\end{figure}

In Fig.~\ref{MCMineINEL900}, comparisons with distributions obtained with the PYTHIA 6 Perugia 0 tune~\cite{Skands:2010ak}, PHOJET~\cite{Bopp:1998rc}, PYTHIA 8 Monash tune~\cite{Sjostrand:2014zea} and EPOS LHC~\cite{Pierog:2013ria} models are shown for INEL events at 0.9 TeV (left plot) and 7 TeV (right plot).
At \hbox{0.9 TeV}, PHOJET, PYTHIA 8 Monash tune and EPOS LHC cannot reproduce the tails, and the lowest values of the multiplicity distributions, while PYTHIA 6 Perugia 0 tune does not reproduce the data at all. 
At 7 TeV, both PHOJET and PYTHIA 6 strongly underestimate the tails of the multiplicity distributions. PYTHIA 8, with the Monash tune that uses LHC data, reproduces the tails for the wider pseudorapidity range, but shows an enhancement in the peak region. EPOS LHC models the distributions well, both in the first bins, which are dominated by diffractive events, and in the tails. 

The evolution of the multiplicity distributions with the center-of-mass energy $\sqrt{s}$ can be studied using the KNO variable $N_{\text{ch}}/\langle N_{\text{ch}}\rangle$~\cite{Koba:1972ng}.
KNO scaling violation is observed if the tails of the distributions increase with increasing energy.
The violation increases when going to larger pseudorapidity ranges.
This behavior was already observed at central rapidities \cite{Adam:2015gka}, and, therefore, it is not investigated any further.

\begin{figure}[htbp]
\centering
\includegraphics[width=\textwidth]{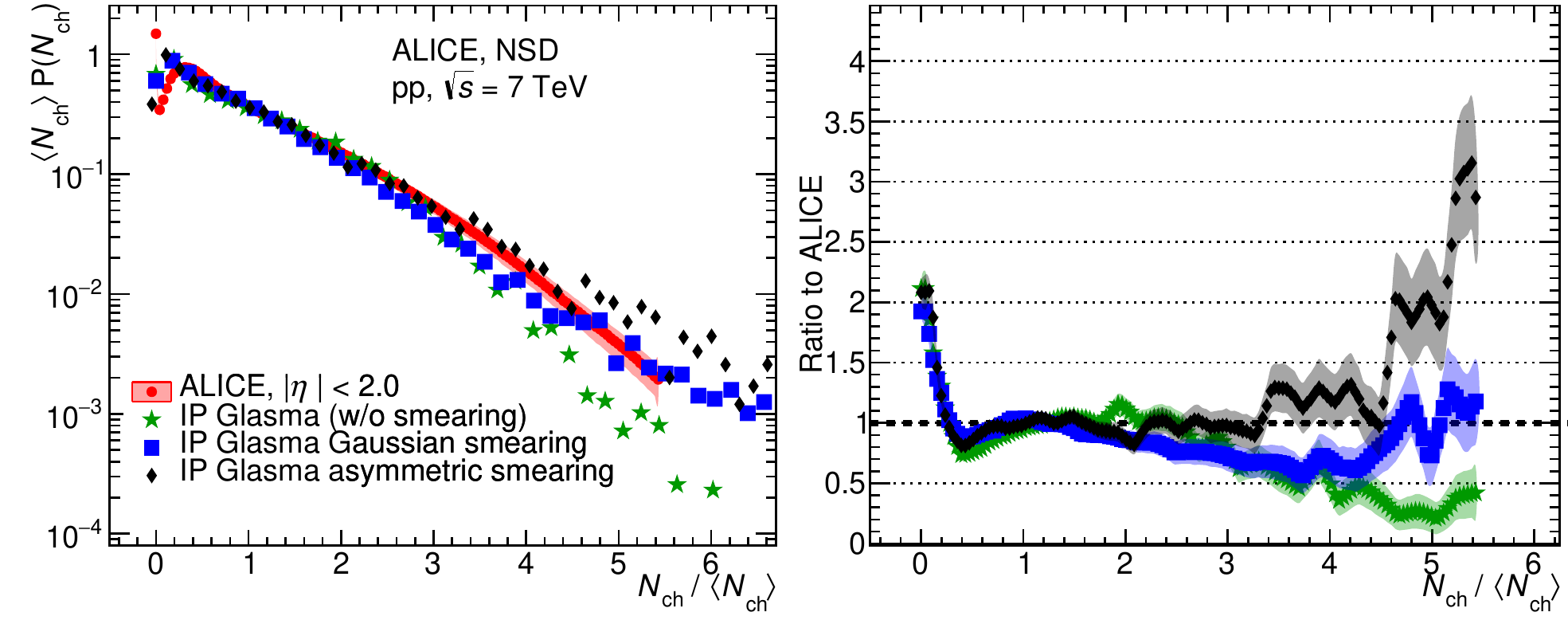}
\caption{Charged-particle multiplicity distributions for pp collisions at $\sqrt{s}=7$ TeV compared to distributions from the IP-Glasma model with the ratio between $Q_{s}$ and the color charge density either fixed (green stars), allowed to fluctuate with a Gaussian (blue squares)~\cite{Schenke:2013dpa} or with additional fluctuations of proton saturation scale (black diamonds)~\cite{McLerran:2015qxa}.}
\label{IPGlasmaMineNSD7000}
\end{figure}

The multiplicity distributions at 7 TeV are compared to those from the IP-Glasma model \cite{Schenke:2013dpa}.
This model is based on the Color Glass Condensate (CGC)~\cite{Iancu:2003xm}.
It has been shown that particle multiplicities are generated following an NBD within the CGC framework~\cite{Gelis:2009wh}.
Moreover, the multiplicity distribution generated by the decay of the Glasma flux tubes~\cite{McLerran:2008es} is a NBD with parameter $k$ (see following Sect.) $\propto Q_{\text{s}}^{2}S_{\perp}$, in which $Q_{\text{s}}$ is the gluon saturation scale and $S_{\perp}$ is the transverse overlap area of the collision~\cite{Schenke:2013dpa}.
The CGC based IP-Glasma model, therefore, has a built-in source of multiplicity fluctuations.
In Fig.~\ref{IPGlasmaMineNSD7000}, the distribution for $\vert\eta\vert<2.0$ is shown together with the IP-Glasma model distributions as a function of the KNO variable $N_{\text{ch}}/\langle N_{\text{ch}}\rangle$. 
The IP-Glasma distribution, shown in green stars, generated with a fixed ratio between $Q_{s}$ and density of color charge, thus introducing no fluctuations. 
The blue squares distribution is generated with fluctuations of the color charge density around the mean value following a Gaussian distribution with a width of $\sigma=0.09$. The black diamonds distribution includes an additional source of fluctuations, dominantly of non-perturbative origin, from stochastic splitting of color dipoles that is not accounted for in the conventional frameworks of CGC~\cite{McLerran:2015qxa}. In the IP-Glasma model shown, the evolution of color charges in the rapidity direction still needs to be implemented and in the present model the low multiplicity bins are not reproduced for the wide pseudorapidity range presented.

\subsection{Double NBD fits}
\label{Sec:nbd}
Already at 0.9 TeV for $\vert\eta\vert<1.3$ in NSD events, ALICE observed that the distributions are not well described by a single NBD parameterization~\cite{Aamodt:2010ft}.
For this reason, a parameterization with the sum of two NBDs has been performed for the NSD, INEL and INEL$>0$ event samples.
The fits are plotted together with the results in Figs. \ref{V0AND900}-\ref{V0AND8000}, and parameters are given in Tab.~\ref{tab:NBDNSD} for the NSD event sample, in Tab.~\ref{tab:NBDINEL} for the INEL event sample and in Tab.~\ref{tab:NBDINELGT0} for the INEL$>$0 event sample. 

\begin{table}[htb] \centering
\tabulinesep=1mm
\tabcolsep=0.11cm
\footnotesize
   \begin{tabu}{c l l l l l l l } \hline
  $\sqrt{s}$ (TeV)& $\eta$ Range & $\lambda$ & $\alpha$ & $\langle n\rangle_{\text{1}}$ & $k_{\text{1}}$ & $\langle n\rangle_{\text{2}}$ & $k_{\text{2}}$\\ \hline\hline
         &  								 & 0.95 & 0.49 &  10.16 & 3.60 & 24.86 & 3.83 \\
         \rowcolor{LightCyan}
  \cellcolor{White}		& \cellcolor{White}$\vert\eta\vert<2.0$  & 0.94 $\pm$ 0.01 & 0.52 $\pm$ 0.05 &  10.28 $\pm$ 0.69 & 3.49 $\pm$ 0.30 & 25.08 $\pm$ 1.96 & 3.96 $\pm$ 0.63 \\
 		& 								  & 0.94 & 0.55 &  10.42 & 3.38 & 25.37 & 4.13 \\ \cline{2-8}
 		& 								  & 0.95 & 0.38  & 11.76 & 4.12 & 28.00 & 3.27   \\
 		\rowcolor{LightCyan}
	\cellcolor{White} 	&  \cellcolor{White}$\vert\eta\vert<2.4$  & 0.95 $\pm$ 0.01 & 0.27 $\pm$ 0.12  & 10.94 $\pm$ 1.14 & 4.76 $\pm$ 2.91 & 25.35 $\pm$ 4.40 & 2.80 $\pm$ 0.96   \\ 
	   & 								  & 0.95 & 0.09  & 10.75 & 25.98 & 21.79 & 2.20   \\  \cline{2-8}
	  	& 								  & 0.94 & 0.27 &  14.41 & 5.85 & 33.00 & 2.99   \\
	  	\rowcolor{LightCyan}
	\cellcolor{White} 0.9    &  \cellcolor{White}$\vert\eta\vert<3.0$  & 0.94 $\pm$ 0.01 & 0.17 $\pm$ 0.06 &  13.68 $\pm$ 0.32 & 9.81 $\pm$ 9.82 & 29.62 $\pm$ 2.85 & 2.54 $\pm$ 0.51   \\
	        & 								  & 0.95 & 0.11 &  13.48 & 21.75 & 27.58 & 2.35  \\ \cline{2-8}
	     & 								  & 0.94 & 0.37 & 17.09 & 4.74 & 38.94 & 3.43 \\
	     \rowcolor{LightCyan}
\cellcolor{White}	       &  \cellcolor{White}$\vert\eta\vert<3.4$  & 0.94 $\pm$ 0.01 & 0.26 $\pm$ 0.09 & 15.80 $\pm$ 0.95 & 5.95 $\pm$ 2.96 & 35.26 $\pm$ 4.35 & 2.98 $\pm$ 0.77 \\
	     & 								  & 0.94 & 0.20 & 14.99 & 18.98 & 32.83 & 2.78 \\ \cline{2-8}
	   & 							     & 0.95 & 0.42 & 20.31 & 5.17 & 46.55 & 4.21  \\
	   \rowcolor{LightCyan}
	\cellcolor{White}	&  \cellcolor{White}$-3.4<\eta<+5.0$     & 0.95 $\pm$ 0.01 & 0.36 $\pm$ 0.07 & 19.27 $\pm$ 1.45 & 5.48 $\pm$ 1.18 & 43.41 $\pm$ 4.93 & 3.85 $\pm$ 1.03  \\ 
		 & 							     & 0.94 & 0.16 & 18.09 & 15.32 & 36.64 & 2.91  \\ \hline\hline
		& 								  & 0.96	& 0.36 & 10.72 & 2.92 & 37.66 & 2.53  \\
		\rowcolor{LightCyan}
  \cellcolor{White}			& \cellcolor{White} $\vert\eta\vert<2.0$  & 0.94 $\pm$ 0.01	& 0.37 $\pm$ 0.02 & 10.63 $\pm$ 0.43 & 2.91 $\pm$ 0.22 & 36.84 $\pm$ 1.41 & 2.51 $\pm$ 0.20  \\
            & 								  & 0.93	& 0.38 & 10.56 & 2.89 & 36.05 & 2.51  \\ \cline{2-8}
            & 									  &  0.97 & 0.36 & 13.19 & 3.13 & 45.86 & 2.61 \\ 
            \rowcolor{LightCyan}
   \cellcolor{White}      	& \cellcolor{White} $\vert\eta\vert<2.4$  &  0.95 $\pm$ 0.01& 0.35 $\pm$ 0.03 & 12.73 $\pm$ 0.68 & 3.17 $\pm$ 0.40 & 43.05 $\pm$ 2.21 & 2.47 $\pm$ 0.27 \\ 
         	& 									  &  0.92 & 0.34 & 12.25 & 3.25 & 40.12 & 2.36 \\ \cline{2-8}
         	& 								  & 0.97	& 0.36 & 16.73 & 3.22 & 57.56 & 2.75  \\ 
         	\rowcolor{LightCyan}
   \cellcolor{White}     7 & \cellcolor{White} $\vert\eta\vert<3.0$  & 0.94 $\pm$ 0.01	& 0.32 $\pm$ 0.03 & 15.55 $\pm$ 0.94 & 3.52 $\pm$ 0.58 & 52.08 $\pm$ 2.93 & 2.45 $\pm$ 0.29  \\ 
           & 									  & 0.91 & 0.26 & 14.28 & 4.36 & 45.78 & 2.15  \\ \cline{2-8}
             & 								  & 0.97 	& 0.38 & 19.45 & 3.15 & 65.72  & 2.89 \\
             \rowcolor{LightCyan}
     \cellcolor{White}        & \cellcolor{White} $\vert\eta\vert<3.4$  & 0.94 $\pm$ 0.01 	& 0.31 $\pm$ 0.03 & 17.43 $\pm$ 1.05 & 3.68 $\pm$ 0.65 & 57.38 $\pm$ 3.33  & 2.43 $\pm$ 0.30 \\
             & 								  & 0.91 	& 0.21 & 15.48 & 5.83 & 48.46  & 2.02 \\ \cline{2-8}
             & 							     &  0.98 & 0.37 & 23.17 & 3.52 & 77.02 & 3.11 \\
             \rowcolor{LightCyan}
 	\cellcolor{White}		& \cellcolor{White} $-3.4<\eta<+5.0$     &  0.94 $\pm$ 0.01 & 0.30 $\pm$ 0.03 & 20.74 $\pm$ 1.28 & 4.18 $\pm$ 0.77 & 66.40 $\pm$ 4.08 & 2.54 $\pm$ 0.34 \\ 
 			&								     &  0.90 & 0.16 & 18.52 & 9.63 & 53.51 & 1.97 \\ \hline \hline
 			& 								  & 0.96	 & 0.44 & 12.58 & 2.38 & 42.18 & 2.97 \\
 			\rowcolor{LightCyan}
  	\cellcolor{White}		&  \cellcolor{White}$\vert\eta\vert<2.0$  & 0.94 $\pm$ 0.01	 & 0.45 $\pm$ 0.03 & 12.37 $\pm$ 0.79 & 2.38 $\pm$ 0.15 & 41.16 $\pm$ 2.01 & 2.93 $\pm$ 0.29 \\
  			& 								  & 0.93	 & 0.46 & 12.21 & 2.38 & 40.20 & 2.91 \\ \cline{2-8}
  			& 								  & 0.96	& 0.37 & 14.18 & 2.77 & 48.59 & 2.70 \\
  			\rowcolor{LightCyan}
    \cellcolor{White}     	& \cellcolor{White} $\vert\eta\vert<2.4$  &  0.94 $\pm$ 0.01 & 0.37 $\pm$ 0.04 & 13.71 $\pm$ 1.10 & 2.75 $\pm$ 0.34 & 45.73 $\pm$ 3.15 & 2.56 $\pm$ 0.37 \\ 
         	& 								  & 0.91	& 0.36 & 13.20 & 2.76 & 42.69 & 2.47 \\ \cline{2-8}
         	& 								  & 0.97	& 0.31 & 16.77 & 3.22 & 57.63 & 2.49 \\
         	\rowcolor{LightCyan}
  \cellcolor{White}      8 & \cellcolor{White} $\vert\eta\vert<3.0$  &  0.94 $\pm$ 0.01	& 0.26 $\pm$ 0.03 & 15.50 $\pm$ 0.99 & 3.63 $\pm$ 0.78 & 51.58 $\pm$ 3.25 & 2.19 $\pm$ 0.29  \\ 
            & 								  & 0.90	& 0.19 & 14.20 & 5.01 & 44.90 & 2.92 \\ \cline{2-8}
            & 								  & 0.97	& 0.33 & 19.54 & 3.21 & 66.21 & 2.64 \\
            \rowcolor{LightCyan}
 \cellcolor{White}            &  \cellcolor{White} $\vert\eta\vert<3.4$  &  0.93 $\pm$ 0.01	& 0.26 $\pm$ 0.03 & 17.57 $\pm$ 0.91 & 3.84 $\pm$ 0.75 & 57.55 $\pm$ 3.17 & 2.22 $\pm$ 0.27 \\
             &								  & 0.90	& 0.18 & 15.93 & 5.73 & 49.51 & 1.95 \\ \cline{2-8}
             & 								  & 0.99	& 0.26 & 21.79 & 4.33 & 73.60 & 2.47 \\
             \rowcolor{LightCyan}
 	\cellcolor{White}		& \cellcolor{White} $-3.4<\eta<+5.0$     & 0.94 $\pm$ 0.01	& 0.17 $\pm$ 0.03 & 20.11 $\pm$ 0.56 & 6.65 $\pm$ 2.45 & 62.60 $\pm$ 3.03 & 2.00 $\pm$ 0.21 \\ 
 			& 								 & 0.90	& 0.14 & 19.22 & 9.96 & 55.01 & 1.92 \\ \hline	
\end{tabu} 
\caption{Double NBD fit parameters for multiplicity distributions, NSD events. The central values of the parameters are printed in the grey-shaded row with their relative uncertainty, while the top row shows the upper bound fit parameters, and the bottom row the lower bound ones. }\label{tab:NBDNSD}
\end{table}
\begin{table}[htb] \centering
\tabulinesep=1mm
\tabcolsep=0.11cm
\footnotesize
   \begin{tabu}{l l l l l l l l  } \hline
  $\sqrt{s}$ (TeV)& $\eta$ Range & $\lambda$ & $\alpha$ & $\langle n\rangle_{\text{1}}$ & $k_{\text{1}}$ & $\langle n\rangle_{\text{2}}$ & $k_{\text{2}}$  \\ \hline\hline
            & 								  & 0.81	&  0.61 & 10.25 & 2.77 & 26.59 & 4.24 \\
            \rowcolor{LightCyan}
 	\cellcolor{White}		&  \cellcolor{White} $\vert\eta\vert<2.0$  & 0.81 $\pm$ 0.01	&  0.64 $\pm$ 0.05 & 10.44 $\pm$ 0.97 & 2.69 $\pm$ 0.20 & 27.03 $\pm$ 2.71 & 4.45 $\pm$ 0.91 \\
 			 &								  & 0.81	&  0.68 & 10.67 & 2.61 & 27.62 & 4.73 \\ \cline{2-8}
 			& 								  &  0.80	& 0.39 & 10.72 & 3.75 & 27.34 & 3.07 \\
 			\rowcolor{LightCyan}
     \cellcolor{White}    	& \cellcolor{White} $\vert\eta\vert<2.4$  &  0.81 $\pm$ 0.01	& 0.33 $\pm$ 0.05 & 10.12 $\pm$ 0.61 & 4.00 $\pm$ 0.93 & 25.48 $\pm$ 2.11 & 2.80 $\pm$ 0.47 \\ 
         	& 								  &  0.81 & 0.26 & 9.51 & 4.72 & 23.52 & 2.52 \\  \cline{2-8}
         	& 								  & 0.80	& 0.38 & 13.81 & 4.14 & 34.33 & 3.20   \\
         	\rowcolor{LightCyan}
 \cellcolor{White}       0.9 & \cellcolor{White} $\vert\eta\vert<3.0$  & 0.80 $\pm$ 0.01	& 0.31 $\pm$ 0.05 & 12.91 $\pm$ 0.66 & 4.83 $\pm$ 1.39 & 31.45 $\pm$ 2.67 & 2.83 $\pm$ 0.50   \\ 
           & 								  & 0.80	& 0.24 & 12.23 & 6.29 & 28.96 & 2.56   \\ \cline{2-8}
            & 								  &  0.80	& 0.46 & 16.52 & 3.73 & 40.48 & 3.67 \\
      \rowcolor{LightCyan}
      \cellcolor{White}       & \cellcolor{White} $\vert\eta\vert<3.4$  &  0.80 $\pm$ 0.01	& 0.40 $\pm$ 0.05 & 15.30 $\pm$ 0.95 & 4.10 $\pm$ 0.63 & 37.39 $\pm$ 2.97 & 3.33 $\pm$ 0.56 \\
             & 								  &  0.80	& 0.36 & 14.46 & 4.54 & 35.52 & 3.24 \\ \cline{2-8}
             & 								  & 0.80	& 0.50 & 19.87 & 4.20 & 48.17 & 4.49  \\
             \rowcolor{LightCyan}
 	\cellcolor{White}		& \cellcolor{White} $-3.4<\eta<+5.0$     & 0.80 $\pm$ 0.01	& 0.46 $\pm$ 0.04 & 18.83 $\pm$ 1.08 & 4.32 $\pm$ 0.45 & 45.16 $\pm$ 3.18 & 4.16 $\pm$ 0.69  \\
 			& 								  & 0.80	& 0.43 & 17.86 & 4.53 & 42.56 & 3.97  \\  \hline\hline
 			& 								 & 0.80	& 0.42 & 10.51 & 2.44 & 38.31 & 2.64 \\
 			\rowcolor{LightCyan}
  	\cellcolor{White}		& \cellcolor{White} $\vert\eta\vert<2.0$  & 0.79 $\pm$ 0.01	& 0.43 $\pm$ 0.02 & 10.42 $\pm$ 0.49 & 2.43 $\pm$ 0.17& 37.54 $\pm$ 1.56 & 2.63 $\pm$ 0.23 \\
  			& 								 & 0.78	& 0.44 & 10.35 & 2.41 & 36.81 & 2.64 \\  \cline{2-8}
  			& 								  & 0.80	& 0.39 & 12.61 & 2.77 & 45.69 & 2.63   \\ 
  			\rowcolor{LightCyan}
     \cellcolor{White}    	& \cellcolor{White} $\vert\eta\vert<2.4$  & 0.79 $\pm$ 0.01	& 0.38 $\pm$ 0.03 & 12.12 $\pm$ 0.71 & 2.79 $\pm$ 0.38 & 42.87 $\pm$ 2.36 & 2.48 $\pm$ 0.28   \\ 
         	& 								  & 0.77	& 0.37 & 11.62 & 2.85 & 39.92 & 2.37   \\  \cline{2-8}	
         	& 							  & 0.80	& 0.40 & 16.17 & 2.77 & 57.70 & 2.82 \\ 
         	\rowcolor{LightCyan}
   \cellcolor{White}     7 & \cellcolor{White} $\vert\eta\vert<3.0$  & 0.78 $\pm$ 0.01	& 0.35 $\pm$ 0.03 & 14.75 $\pm$ 1.01 & 3.13 $\pm$ 0.63 & 51.55 $\pm$ 3.28 & 2.45 $\pm$ 0.32 \\ 
            & 							  & 0.75	& 0.28 & 13.38 & 4.47 & 44.74 & 2.11 \\   \cline{2-8}	
            & 								  & 0.80 	& 0.35 & 19.04 & 2.67 & 66.42 & 3.03  \\
            \rowcolor{LightCyan}
     \cellcolor{White}        & \cellcolor{White} $\vert\eta\vert<3.4$  & 0.78 $\pm$ 0.01 	& 0.35 $\pm$ 0.03 & 16.69 $\pm$ 1.14 & 3.15 $\pm$ 0.57 & 57.31 $\pm$ 3.63 & 2.48 $\pm$ 0.33  \\
             & 								  & 0.75 	& 0.26 & 14.82 & 4.47 & 48.93 & 2.11  \\  \cline{2-8}	
             & 							     & 0.81	& 0.41 & 22.34 & 3.03 & 77.02 & 3.20  \\
             \rowcolor{LightCyan}
 		\cellcolor{White}	& \cellcolor{White} $-3.4<\eta<+5.0$     & 0.78 $\pm$ 0.01	& 0.33 $\pm$ 0.03 & 19.54 $\pm$ 1.21 & 3.75 $\pm$ 0.77 & 65.47 $\pm$ 4.01 & 2.53 $\pm$ 0.33  \\ 
 			& 							     & 0.75	& 0.24 & 17.57 & 5.70 & 54.79 & 2.12  \\  \hline	 \hline	
 			& 							  & 0.81	& 0.51 & 11.13 & 1.89 & 41.16 & 3.04 \\
 			\rowcolor{LightCyan}
  	\cellcolor{White}		& \cellcolor{White} $\vert\eta\vert<2.0$  & 0.80 $\pm$ 0.01	& 0.52 $\pm$ 0.03 & 10.87 $\pm$ 0.84 & 1.93 $\pm$ 0.25 & 40.01 $\pm$ 2.39 & 2.98 $\pm$ 0.35 \\
  			& 							  & 0.79	& 0.52 & 10.67 & 1.96 & 38.95 & 2.94 \\ \cline{2-8}
  			& 								  & 0.80	& 0.43 & 12.58 & 2.38 & 47.19 & 2.73 \\
  			\rowcolor{LightCyan}
     \cellcolor{White}    	& \cellcolor{White} $\vert\eta\vert<2.4$  & 0.79 $\pm$ 0.01	& 0.42 $\pm$ 0.04 & 12.08 $\pm$ 0.98 & 2.37 $\pm$ 0.38 & 44.39 $\pm$ 3.12 & 2.58 $\pm$ 0.37  \\ 
         	& 								  & 0.77	& 0.43 & 11.54 & 2.39 & 41.38 & 2.47 \\ \cline{2-8}
         	 & 							  & 0.80	& 0.37 & 15.25 & 2.72 & 56.72 & 2.57 \\
         	 \rowcolor{LightCyan}
   \cellcolor{White}     8 & \cellcolor{White} $\vert\eta\vert<3.0$  & 0.78 $\pm$ 0.01	& 0.32 $\pm$ 0.03 & 13.94 $\pm$ 0.85 & 3.10 $\pm$ 0.64 & 50.50 $\pm$ 3.16 & 2.23 $\pm$ 0.29  \\ 
             & 							  & 0.75	& 0.26 & 12.78 & 4.06 & 43.94 & 1.94 \\ \cline{2-8}	
        & 									  & 0.80	& 0.41 & 18.01 & 2.65 & 65.78 & 2.78  \\
        \rowcolor{LightCyan}
      \cellcolor{White}       & \cellcolor{White} $\vert\eta\vert<3.4$  & 0.78 $\pm$ 0.01	& 0.34 $\pm$ 0.03 & 16.03 $\pm$ 0.86 & 3.08 $\pm$ 0.49 & 57.14 $\pm$ 3.15 & 2.32 $\pm$ 0.28  \\
            & 							    & 0.75	& 0.27 & 14.56 & 3.98 & 49.55 & 2.05  \\ \cline{2-8}
             &							     & 0.82	& 0.37 & 20.95 & 3.16 & 75.52 & 2.78	\\
             \rowcolor{LightCyan}
 	\cellcolor{White}		& \cellcolor{White} $-3.4<\eta<+5.0$     & 0.78 $\pm$ 0.01	& 0.30 $\pm$ 0.03 & 19.03 $\pm$ 0.86 & 3.78 $\pm$ 0.74 & 65.08 $\pm$ 3.74 & 2.29 $\pm$ 0.29	\\ 
 			&							     & 0.75	& 0.23 & 17.61 & 5.20 & 55.11 & 1.99	\\ \hline	
\end{tabu} 
\caption{Double NBD fit parameters for multiplicity distributions, INEL events. The central values of the parameters are printed in the grey-shaded row with their relative uncertainty, while the top row shows the upper bound fit parameters, and the bottom row the lower bound ones.} \label{tab:NBDINEL}
\end{table}

\begin{table}[htb] \centering
\tabulinesep=1mm
\tabcolsep=0.11cm
\footnotesize
   \begin{tabu}{l l l l l l l l  } \hline
  $\sqrt{s}$ (TeV)& $\eta$ Range & $\lambda$ & $\alpha$ & $\langle n\rangle_{\text{1}}$ & $k_{\text{1}}$ & $\langle n\rangle_{\text{2}}$ & $k_{\text{2}}$ \\ \hline\hline
 			& 								  & 1.00	& 0.53 & 10.43 & 3.40 & 25.55 & 3.94 \\
 			\rowcolor{LightCyan}
 	\cellcolor{White} 		& \cellcolor{White}  $\vert\eta\vert<2.0$  & 1.00 $\pm$ 0.01	& 0.56 $\pm$ 0.04 & 10.54 $\pm$ 0.69 & 3.31 $\pm$ 0.21 & 25.78 $\pm$ 2.04 & 4.08 $\pm$ 0.65 \\
            & 								  & 1.00	& 0.59 & 10.67 & 3.23 & 26.00 & 4.26 \\ \cline{2-8}
			& 								 & 1.00	& 0.49 & 12.35 & 3.40 & 30.27 & 3.75 \\
			\rowcolor{LightCyan}
    \cellcolor{White}      	& \cellcolor{White}  $\vert\eta\vert<2.4$  &  1.00 $\pm$ 0.01	& 0.45 $\pm$ 0.06 & 11.84 $\pm$ 1.11 & 3.37 $\pm$ 0.37 & 28.64 $\pm$ 2.93 & 3.49 $\pm$ 0.73 \\ 
         	& 								  & 0.99	& 0.11 & 10.56 & 19.63 & 22.06 & 2.23 \\  \cline{2-8}
         	& 								  & 0.97	& 0.32 & 14.35 & 4.81 & 33.95 & 3.13 \\
         	\rowcolor{LightCyan}
\cellcolor{White}         0.9 & \cellcolor{White}  $\vert\eta\vert<3.0$  & 0.97 $\pm$ 0.01	& 0.23 $\pm$ 0.07 & 13.42 $\pm$ 0.56 & 6.30 $\pm$ 3.33 & 30.77 $\pm$ 3.00 & 2.71 $\pm$ 0.55   \\ 
            & 								  & 0.97	& 0.14 & 12.99 & 12.54 & 28.07 & 2.41 \\ \cline{2-8}
             & 								  & 0.95	& 0.40 & 16.98 & 4.26 & 39.78 & 3.55 \\
             \rowcolor{LightCyan}
\cellcolor{White}              & \cellcolor{White}  $\vert\eta\vert<3.4$  &  0.95 $\pm$ 0.01	& 0.33 $\pm$ 0.05 & 15.76 $\pm$ 0.90 & 4.80 $\pm$ 0.98 & 36.74 $\pm$ 2.96 & 3.22 $\pm$ 0.55 \\
            & 								  & 0.95	& 0.29 & 14.88 & 5.50 & 34.84 & 3.11 \\ \cline{2-8}
            & 								  & 0.93	& 0.46 & 20.45 & 4.61 & 47.77 & 4.41 \\
            \rowcolor{LightCyan}
\cellcolor{White}  			& \cellcolor{White}  $-3.4<\eta<+5.0$     & 0.92 $\pm$ 0.01	& 0.41 $\pm$ 0.05 & 19.37 $\pm$ 1.10 & 4.77 $\pm$ 0.54 & 44.72 $\pm$ 3.20 & 4.07 $\pm$ 0.68   \\ 
 			& 								  & 0.92	& 0.37 & 18.35 & 5.06 & 42.06 & 3.87 \\ \hline \hline
 			& 								  & 1.02	& 0.37 & 10.96 & 3.04 & 37.57 & 2.55 \\
 			\rowcolor{LightCyan}
 \cellcolor{White}  			& \cellcolor{White}  $\vert\eta\vert<2.0$  & 1.00 $\pm$ 0.01	& 0.37 $\pm$ 0.02 & 10.88 $\pm$ 0.42 & 3.03 $\pm$ 0.22 & 36.77 $\pm$ 1.41 & 2.53 $\pm$ 0.21 \\
  			& 								  & 0.99	& 0.38 & 10.81 & 3.01 & 36.01 & 2.53 \\ \cline{2-8}	
  			& 								  & 1.02	& 0.36 & 13.31 & 3.18 & 45.58 & 2.62 \\
  			\rowcolor{LightCyan}
\cellcolor{White}          	& \cellcolor{White}  $\vert\eta\vert<2.4$  & 1.00 $\pm$ 0.01	& 0.35 $\pm$ 0.03 & 12.87 $\pm$ 0.69 & 3.21 $\pm$ 0.41 & 42.83 $\pm$ 2.27 & 2.48 $\pm$ 0.27   \\ 
         	& 								  & 0.97	& 0.34 & 12.41 & 3.28 & 39.97 & 2.38 \\ \cline{2-8}	
         	& 								  & 1.01	& 0.36 & 16.88 & 3.19 & 57.27 & 2.78 \\
         	\rowcolor{LightCyan}
\cellcolor{White}         7 & \cellcolor{White}  $\vert\eta\vert<3.0$  & 0.98 $\pm$ 0.01	& 0.32 $\pm$ 0.03 & 15.58 $\pm$ 1.00 & 3.56 $\pm$ 0.67 & 51.50 $\pm$ 3.09 & 2.44 $\pm$ 0.30 \\ 
             & 								  & 0.94	& 0.24 & 14.14 & 4.81 & 44.68 & 2.10 \\ \cline{2-8}	
             & 								  & 0.99	& 0.39 & 19.79 & 3.06 & 65.81 & 2.97 \\
             \rowcolor{LightCyan}
 \cellcolor{White}             & \cellcolor{White}  $\vert\eta\vert<3.4$  & 0.96 $\pm$ 0.01	& 0.31 $\pm$ 0.03 & 17.53 $\pm$ 1.15 & 3.63 $\pm$ 0.69 & 57.02 $\pm$ 3.52 & 2.45 $\pm$ 0.32  \\
             & 								  & 0.93	& 0.20 & 15.45 & 6.03 & 48.00 & 2.01 \\ \cline{2-8}	
             & 								  & 0.98	& 0.39 & 23.40 & 3.31 & 77.14 & 3.21 \\
             \rowcolor{LightCyan}
 \cellcolor{White} 			& \cellcolor{White}  $-3.4<\eta<+5.0$     & 0.94 $\pm$ 0.01	& 0.31 $\pm$ 0.03 & 20.70 $\pm$ 1.37 & 3.94 $\pm$ 0.72 & 66.27 $\pm$ 4.07 & 2.60 $\pm$ 0.34  \\ 
 			& 								  & 0.90	& 0.17 & 18.24 & 8.39 & 53.51 & 2.02 \\ \hline	\hline	
 			& 							  & 1.01	& 0.43 & 10.99 & 2.54 & 39.09 & 2.77  \\
 			\rowcolor{LightCyan}
 \cellcolor{White}  			& \cellcolor{White}  $\vert\eta\vert<2.0$  & 1.00 $\pm$ 0.01	& 0.43 $\pm$ 0.02 & 10.83 $\pm$ 0.52 & 2.54 $\pm$ 0.18 & 38.14 $\pm$ 1.48 & 2.73 $\pm$ 0.21  \\
  			& 							  & 0.99	& 0.44 & 10.70 & 2.55 & 37.24 & 2.71  \\ \cline{2-8}	
  			& 							  & 1.01	& 0.38 & 12.92 & 2.91 & 46.16 & 2.61  \\
  			\rowcolor{LightCyan}
\cellcolor{White}          	& \cellcolor{White}  $\vert\eta\vert<2.4$  & 0.99 $\pm$ 0.01	& 0.37 $\pm$ 0.03 & 12.45 $\pm$ 0.81 & 2.93 $\pm$ 0.39 & 43.33 $\pm$ 2.62 & 2.47 $\pm$ 0.31  \\ 
         	& 							  & 0.96	& 0.35 & 11.94 & 2.99 & 40.32 & 2.36  \\ \cline{2-8}	
         	& 							  & 1.00	& 0.33 & 15.78 & 3.24 & 55.92 & 2.50  \\
         	\rowcolor{LightCyan}
\cellcolor{White}         8 & \cellcolor{White}  $\vert\eta\vert<3.0$  & 0.97 $\pm$ 0.01	& 0.28 $\pm$ 0.03 & 14.61 $\pm$ 0.77 & 3.73 $\pm$ 0.80 & 50.01 $\pm$ 2.90 & 2.19 $\pm$ 0.27   \\ 
            & 							  & 0.94	& 0.22 & 13.55 & 5.02 & 43.78 & 1.93  \\ \cline{2-8}
             & 							  & 0.99	& 0.37 & 18.60 & 3.08 & 64.94 & 2.71  \\
             \rowcolor{LightCyan}
 \cellcolor{White}             & \cellcolor{White}  $\vert\eta\vert<3.4$  & 0.95 $\pm$ 0.01	& 0.30 $\pm$ 0.03 & 16.79 $\pm$ 0.80 & 3.61 $\pm$ 0.60 & 56.72 $\pm$ 2.97 & 2.28 $\pm$ 0.26  \\
             & 							  & 0.92	& 0.23 & 15.34 & 4.83 & 49.25 & 2.03  \\ \cline{2-8}
             & 							  & 0.98	& 0.34 & 21.85 & 3.47 & 75.56 & 2.79  \\
             \rowcolor{LightCyan}
\cellcolor{White}  			& \cellcolor{White}  $-3.4<\eta<+5.0$     & 0.93 $\pm$ 0.01	& 0.28 $\pm$ 0.03 & 20.00 $\pm$ 0.92 & 4.08 $\pm$ 0.77 & 65.51 $\pm$ 3.77 & 2.32 $\pm$ 0.30	 \\ 
 			& 							  & 0.89	& 0.19 & 18.43 & 6.23 & 54.67 & 1.96  \\ \hline	
\end{tabu} 
\caption{Double NBD fit parameters for multiplicity distributions, INEL$>$0 events. The central values of the parameters are printed in the grey-shaded row with their relative uncertainty, while the top row shows the upper bound fit parameters, and the bottom row the lower bound ones.} \label{tab:NBDINELGT0}
\end{table}

The distributions have been fitted using the function
\begin{equation}
\text{P}(n)=\lambda[\alpha \text{P}_{\text{NBD}}(n,\langle n\rangle_{\text{1}},k_{\text{1}})+(1-\alpha)\text{P}_{\text{NBD}}(n,\langle n\rangle_{\text{2}},k_{\text{2}})]\,,
\end{equation}
where
\begin{equation}
\text{P}_{\text{NBD}}(n,\langle n\rangle,k)=\frac{\Gamma(n+k)}{\Gamma(k)\Gamma(n+1)}\bigg(\frac{\langle n\rangle}{k+\langle n\rangle}\bigg)^{n}\bigg(\frac{k}{k+\langle n\rangle}\bigg)^{k}\,.
\end{equation}
NBDs do not describe the bin with $N_{ch}=0$ and the first bins for the wider pseudorapidities, in which there is a rise in the number of events measured due to diffractive events (bins removed from the fit).
To account for this a normalization factor $\lambda$ is introduced \cite{Adam:2015gka}.
The systematic uncertainty for the efficiency correction produces a fully correlated shift of the distribution, and therefore is omitted in the fitted data sample. The other sources of uncertainty are kept in the data. From fitting with the above uncertainties included, one obtains the parameter in the grey-shaded row.
The parameters of two extreme cases are printed as well in tables~\ref{tab:NBDNSD}-\ref{tab:NBDINELGT0} to allow the reader to have an estimate on how much the fit parameters change due to the remaining correlations present. The upper and lower case correspond to fitting the distribution obtained adding and subtracting material in the detector.  
The behavior of the fit parameters is consistent with what was observed by CMS \cite{Khachatryan:2010nk,Ghosh:2012xh}.
The average multiplicity of the soft (first) component, $\langle n\rangle_{\text{1}}$, increases with increasing energy and pseudorapidity range.
The parameter $k_{\text{1}}$ increases with increasing pseudorapidity range and decreases with center-of-mass energy.
For the semi-hard (second) component, the $\langle n\rangle_{\text{2}}$ parameter behaves similarly to $\langle n\rangle_{\text{1}}$.
Moreover, it is noted that $\langle n\rangle_{\text{2}}\simeq3\langle n\rangle_{\text{1}}$.
The parameter $\alpha$ reflects the fraction of the soft events.
The percentage of soft events decreases with increasing pseudorapidity range and energy indicating that a higher percentage of semi-hard events is then present.

For KNO scaling to hold, the parameter $k$ must be constant with energy.
This means that KNO scaling is violated for all pseudorapidity ranges for both the soft and semi-hard components in the NSD, INEL and INEL$>$0 event samples.
Three possible scenarios are proposed in the model by Ugoccioni and Giovannini~\cite{Giovannini:1998zb} concerning KNO scaling in the semi-hard component.
Parameter $k_{\text{2}}$ can be constant, decrease linearly with increasing energy showing violation, or decrease with energy asymptotically to a constant value showing weak violation.
The last scenario appears to be in agreement with the values obtained for $k_{\text{2}}$, since the decrease from 0.9 TeV to 7 TeV is very strong, while the values are compatible for 7 TeV and 8 TeV.
The analysis of 13 TeV data will reveal if the values will be compatible also in that case, or if it is only due to the vicinity of the 7 and 8 TeV energies.

\section{Summary and conclusions}
\label{Sec:conc}
Data from the SPD and the FMD were used to access a uniquely wide pseudorapidity coverage at the LHC of more than eight units in pseudorapidity, from $-3.4<\eta<5.0$.
The charged-particle multiplicity distributions obtained from these data were presented for three pp collision energies, $\sqrt{s}=$0.9, 7, and 8 TeV, and for three different event classes, INEL, INEL$>0$, and NSD.
The results shown extend the pseudorapidity coverage of the earlier results published by ALICE and CMS, and, consequently, the high-multiplicity reach.
The extension of pseudorapidity coverage has higher systematic uncertainties due to the unknown fraction of the material budget in front of the FMD, estimated to be up to 14$\%$.

The multiplicity distributions for 7 TeV collisions measured for NSD events are in agreement with those from CMS.
For 0.9 TeV, the results shown are systematically lower in the low-multiplicity bins when compared to the CMS measurement.  
This is most probably due to a different estimation of the diffractive masses distribution used to tune the simulations for the efficiency correction between CMS and ALICE.
At 0.9 TeV, PYTHIA, PHOJET and EPOS LHC cannot reproduce the multiplicity distributions, although PHOJET is closer to the results being tuned using LHC data.
At 7~TeV, PYTHIA 6 and PHOJET strongly underestimate the fraction of high multiplicity events. PYTHIA 8 slightly underestimates the tails of the distributions, while EPOS LHC reproduces both the low and the high multiplicity events, showing better capabilities to model diffraction. 
The Color Glass Condensate based IP-Glasma model produces distributions, which underestimate the fraction of high multiplicity events, but introducing fluctuations in the saturation momentum allows to reproduce the measurements better. 
Double Negative Binomial distributions composed of a soft and hard component are fitted to the measured distributions.
The fraction of soft events decreases with increasing pseudorapidity range and with the increasing collision energy.
KNO scaling is violated for the three considered event classes, at all the collision energies probed.
The 13 TeV data analysis will help delving into the description of the components of the multiplicity distributions.

\clearpage
\newenvironment{acknowledgement}{\relax}{\relax}
\begin{acknowledgement}
\section*{Acknowledgments}

The ALICE Collaboration would like to thank all its engineers and technicians for their invaluable contributions to the construction of the experiment and the CERN accelerator teams for the outstanding performance of the LHC complex.
The ALICE Collaboration gratefully acknowledges the resources and support provided by all Grid centres and the Worldwide LHC Computing Grid (WLCG) collaboration.
The ALICE Collaboration acknowledges the following funding agencies for their support in building and running the ALICE detector:
A. I. Alikhanyan National Science Laboratory (Yerevan Physics Institute) Foundation (ANSL), State Committee of Science and World Federation of Scientists (WFS), Armenia;
Austrian Academy of Sciences and Nationalstiftung f\"{u}r Forschung, Technologie und Entwicklung, Austria;
Ministry of Communications and High Technologies, National Nuclear Research Center, Azerbaijan;
Conselho Nacional de Desenvolvimento Cient\'{\i}fico e Tecnol\'{o}gico (CNPq), Universidade Federal do Rio Grande do Sul (UFRGS), Financiadora de Estudos e Projetos (Finep) and Funda\c{c}\~{a}o de Amparo \`{a} Pesquisa do Estado de S\~{a}o Paulo (FAPESP), Brazil;
Ministry of Science \& Technology of China (MSTC), National Natural Science Foundation of China (NSFC) and Ministry of Education of China (MOEC) , China;
Ministry of Science, Education and Sport and Croatian Science Foundation, Croatia;
Ministry of Education, Youth and Sports of the Czech Republic, Czech Republic;
The Danish Council for Independent Research | Natural Sciences, the Carlsberg Foundation and Danish National Research Foundation (DNRF), Denmark;
Helsinki Institute of Physics (HIP), Finland;
Commissariat \`{a} l'Energie Atomique (CEA) and Institut National de Physique Nucl\'{e}aire et de Physique des Particules (IN2P3) and Centre National de la Recherche Scientifique (CNRS), France;
Bundesministerium f\"{u}r Bildung, Wissenschaft, Forschung und Technologie (BMBF) and GSI Helmholtzzentrum f\"{u}r Schwerionenforschung GmbH, Germany;
General Secretariat for Research and Technology, Ministry of Education, Research and Religions, Greece;
National Research, Development and Innovation Office, Hungary;
Department of Atomic Energy Government of India (DAE) and Council of Scientific and Industrial Research (CSIR), New Delhi, India;
Indonesian Institute of Science, Indonesia;
Centro Fermi - Museo Storico della Fisica e Centro Studi e Ricerche Enrico Fermi and Istituto Nazionale di Fisica Nucleare (INFN), Italy;
Institute for Innovative Science and Technology , Nagasaki Institute of Applied Science (IIST), Japan Society for the Promotion of Science (JSPS) KAKENHI and Japanese Ministry of Education, Culture, Sports, Science and Technology (MEXT), Japan;
Consejo Nacional de Ciencia (CONACYT) y Tecnolog\'{i}a, through Fondo de Cooperaci\'{o}n Internacional en Ciencia y Tecnolog\'{i}a (FONCICYT) and Direcci\'{o}n General de Asuntos del Personal Academico (DGAPA), Mexico;
Nederlandse Organisatie voor Wetenschappelijk Onderzoek (NWO), Netherlands;
The Research Council of Norway, Norway;
Commission on Science and Technology for Sustainable Development in the South (COMSATS), Pakistan;
Pontificia Universidad Cat\'{o}lica del Per\'{u}, Peru;
Ministry of Science and Higher Education and National Science Centre, Poland;
Korea Institute of Science and Technology Information and National Research Foundation of Korea (NRF), Republic of Korea;
Ministry of Education and Scientific Research, Institute of Atomic Physics and Romanian National Agency for Science, Technology and Innovation, Romania;
Joint Institute for Nuclear Research (JINR), Ministry of Education and Science of the Russian Federation and National Research Centre Kurchatov Institute, Russia;
Ministry of Education, Science, Research and Sport of the Slovak Republic, Slovakia;
National Research Foundation of South Africa, South Africa;
Centro de Aplicaciones Tecnol\'{o}gicas y Desarrollo Nuclear (CEADEN), Cubaenerg\'{\i}a, Cuba, Ministerio de Ciencia e Innovacion and Centro de Investigaciones Energ\'{e}ticas, Medioambientales y Tecnol\'{o}gicas (CIEMAT), Spain;
Swedish Research Council (VR) and Knut \& Alice Wallenberg Foundation (KAW), Sweden;
European Organization for Nuclear Research, Switzerland;
National Science and Technology Development Agency (NSDTA), Suranaree University of Technology (SUT) and Office of the Higher Education Commission under NRU project of Thailand, Thailand;
Turkish Atomic Energy Agency (TAEK), Turkey;
National Academy of  Sciences of Ukraine, Ukraine;
Science and Technology Facilities Council (STFC), United Kingdom;
National Science Foundation of the United States of America (NSF) and United States Department of Energy, Office of Nuclear Physics (DOE NP), United States of America.
\end{acknowledgement}
\bibliographystyle{utphys}   
\bibliography{biblio_3}

\providecommand{\href}[2]{#2}\begingroup\raggedright\begin{thebibliography}{10}

\bibitem{ALICE-PUBLIC-2017-005}
{\bfseries ALICE} Collaboration, S.~Acharya {\em et~al.}, ``{The ALICE
  definition of primary particles},'' June, 2017.
\newblock \url{http://cds.cern.ch/record/2270008}.

\bibitem{Aamodt:2010ft}
{\bfseries ALICE} Collaboration, K.~Aamodt {\em et~al.}, ``{Charged-particle
  multiplicity measurement in proton-proton collisions at $\sqrt{s}=0.9$ and
  2.36 TeV with ALICE at LHC},''
  \href{http://dx.doi.org/10.1140/epjc/s10052-010-1339-x}{{\em Eur.Phys.J.}
  {\bfseries C68} (2010) 89},
\href{http://arxiv.org/abs/1004.3034}{{\ttfamily arXiv:1004.3034 [hep-ex]}}.

\bibitem{Aamodt:2010pp}
{\bfseries ALICE} Collaboration, K.~Aamodt {\em et~al.}, ``{Charged-particle
  multiplicity measurement in proton-proton collisions at $\sqrt{s}=7$ TeV with
  ALICE at LHC},'' \href{http://dx.doi.org/10.1140/epjc/s10052-010-1350-2}{{\em
  Eur.Phys.J.} {\bfseries C68} (2010) 345--354},
\href{http://arxiv.org/abs/1004.3514}{{\ttfamily arXiv:1004.3514 [hep-ex]}}.

\bibitem{Adam:2015gka}
{\bfseries ALICE} Collaboration, J.~Adam {\em et~al.}, ``{Charged-particle
  multiplicities in proton-proton collisions at $\sqrt{s} = 0.9$ to 8 TeV},''
  \href{http://dx.doi.org/10.1140/epjc/s10052-016-4571-1}{{\em Eur. Phys. J.}
  {\bfseries C77} no.~1, (2017) 33},
\href{http://arxiv.org/abs/1509.07541}{{\ttfamily arXiv:1509.07541 [nucl-ex]}}.

\bibitem{Khachatryan:2010nk}
{\bfseries CMS} Collaboration, V.~Khachatryan {\em et~al.}, ``{Charged particle
  multiplicities in $pp$ interactions at $\sqrt{s}=0.9$, 2.36, and 7 TeV},''
  \href{http://dx.doi.org/10.1007/JHEP01(2011)079}{{\em JHEP} {\bfseries 1101}
  (2011) 079},
\href{http://arxiv.org/abs/1011.5531}{{\ttfamily arXiv:1011.5531 [hep-ex]}}.

\bibitem{Aad:2010ac}
{\bfseries ATLAS} Collaboration, G.~Aad {\em et~al.}, ``{Charged-particle
  multiplicities in pp interactions measured with the ATLAS detector at the
  LHC},'' \href{http://dx.doi.org/10.1088/1367-2630/13/5/053033}{{\em New J.
  Phys.} {\bfseries 13} (2011) 053033},
\href{http://arxiv.org/abs/1012.5104}{{\ttfamily arXiv:1012.5104 [hep-ex]}}.

\bibitem{Aaij:2014pza}
{\bfseries LHCb} Collaboration, R.~Aaij {\em et~al.}, ``{Measurement of charged
  particle multiplicities and densities in $pp$ collisions at $\sqrt{s}=7\;$TeV
  in the forward region},''
  \href{http://dx.doi.org/10.1140/epjc/s10052-014-2888-1}{{\em Eur. Phys. J.}
  {\bfseries C74} no.~5, (2014) 2888},
\href{http://arxiv.org/abs/1402.4430}{{\ttfamily arXiv:1402.4430 [hep-ex]}}.

\bibitem{Skands:2010ak}
P.~Z. Skands, ``{Tuning Monte Carlo Generators: The Perugia Tunes},''
  \href{http://dx.doi.org/10.1103/PhysRevD.82.074018}{{\em Phys.Rev.}
  {\bfseries D82} (2010) 074018},
\href{http://arxiv.org/abs/1005.3457}{{\ttfamily arXiv:1005.3457 [hep-ph]}}.

\bibitem{Bopp:1998rc}
F.~W. Bopp, R.~Engel, and J.~Ranft, ``{Rapidity gaps and the PHOJET Monte
  Carlo},''
\href{http://arxiv.org/abs/hep-ph/9803437}{{\ttfamily arXiv:hep-ph/9803437
  [hep-ph]}}.

\bibitem{Sjostrand:2014zea}
T.~Sjöstrand, S.~Ask, J.~R. Christiansen, R.~Corke, N.~Desai, P.~Ilten,
  S.~Mrenna, S.~Prestel, C.~O. Rasmussen, and P.~Z. Skands, ``{An Introduction
  to PYTHIA 8.2},'' \href{http://dx.doi.org/10.1016/j.cpc.2015.01.024}{{\em
  Comput. Phys. Commun.} {\bfseries 191} (2015) 159},
\href{http://arxiv.org/abs/1410.3012}{{\ttfamily arXiv:1410.3012 [hep-ph]}}.

\bibitem{Pierog:2013ria}
T.~Pierog, I.~Karpenko, J.~M. Katzy, E.~Yatsenko, and K.~Werner, ``{EPOS LHC:
  Test of collective hadronization with data measured at the CERN Large Hadron
  Collider},'' \href{http://dx.doi.org/10.1103/PhysRevC.92.034906}{{\em Phys.
  Rev.} {\bfseries C92} no.~3, (2015) 034906},
\href{http://arxiv.org/abs/1306.0121}{{\ttfamily arXiv:1306.0121 [hep-ph]}}.

\bibitem{Schenke:2013dpa}
B.~Schenke, P.~Tribedy, and R.~Venugopalan, ``{Multiplicity distributions in
  p+p, p+A and A+A collisions from Yang-Mills dynamics},''
  \href{http://dx.doi.org/10.1103/PhysRevC.89.024901}{{\em Phys.Rev.}
  {\bfseries C89} (2014) 024901},
\href{http://arxiv.org/abs/1311.3636}{{\ttfamily arXiv:1311.3636 [hep-ph]}}.

\bibitem{McLerran:2015qxa}
L.~McLerran and P.~Tribedy, ``{Intrinsic Fluctuations of the Proton Saturation
  Momentum Scale in High Multiplicity p+p Collisions},''
  \href{http://dx.doi.org/10.1016/j.nuclphysa.2015.10.008}{{\em Nucl. Phys.}
  {\bfseries A945} (2016) 216},
\href{http://arxiv.org/abs/1508.03292}{{\ttfamily arXiv:1508.03292 [hep-ph]}}.

\bibitem{Aamodt:2008zz}
{\bfseries ALICE} Collaboration, K.~Aamodt {\em et~al.}, ``{The ALICE
  experiment at the CERN LHC},''
\href{http://dx.doi.org/10.1088/1748-0221/3/08/S08002}{{\em JINST} {\bfseries
  3} (2008) S08002}.

\bibitem{Abbas:2013taa}
{\bfseries ALICE} Collaboration, E.~Abbas {\em et~al.}, ``{Performance of the
  ALICE VZERO system},''
  \href{http://dx.doi.org/10.1088/1748-0221/8/10/P10016}{{\em JINST} {\bfseries
  8} (2013) P10016},
\href{http://arxiv.org/abs/1306.3130}{{\ttfamily arXiv:1306.3130 [nucl-ex]}}.

\bibitem{Cortese:2004aa}
{\bfseries ALICE} Collaboration, P.~Cortese {\em et~al.}, ``{ALICE technical
  design report on forward detectors: FMD, T0 and V0},''
{\em CERN-LHCC-2004-025} (2004) .

\bibitem{Abbas:2013bpa}
{\bfseries ALICE} Collaboration, E.~Abbas {\em et~al.}, ``{Centrality
  dependence of the pseudorapidity density distribution for charged particles
  in Pb-Pb collisions at $\sqrt{s_{\rm NN}}$ = 2.76 TeV},''
  \href{http://dx.doi.org/10.1016/j.physletb.2013.09.022}{{\em Phys. Lett.}
  {\bfseries B726} (2013) 610--622},
\href{http://arxiv.org/abs/1304.0347}{{\ttfamily arXiv:1304.0347 [nucl-ex]}}.

\bibitem{Adam:2015kda}
{\bfseries ALICE} Collaboration, J.~Adam {\em et~al.}, ``{Centrality evolution
  of the charged-particle pseudorapidity density over a broad pseudorapidity
  range in Pb-Pb collisions at $\sqrt{s_{\rm NN}}$ = 2.76 TeV},''
\href{http://arxiv.org/abs/1509.07299}{{\ttfamily arXiv:1509.07299 [nucl-ex]}}.

\bibitem{Abelev:2014ffa}
{\bfseries ALICE} Collaboration, B.~B. Abelev {\em et~al.}, ``{Performance of
  the ALICE Experiment at the CERN LHC},''
  \href{http://dx.doi.org/10.1142/S0217751X14300440}{{\em Int.J.Mod.Phys.}
  {\bfseries A29} (2014) 1430044},
\href{http://arxiv.org/abs/1402.4476}{{\ttfamily arXiv:1402.4476 [nucl-ex]}}.

\bibitem{Collins:1977jy}
P.~D.~B. Collins, {\em {An Introduction to Regge Theory and High-Energy
  Physics}}.
\newblock Cambridge Monographs on Mathematical Physics. Cambridge Univ. Press,
  Cambridge, UK, 2009.
\newblock
\url{http://www-spires.fnal.gov/spires/find/books/www?cl=QC793.3.R4C695}.
\newblock

\bibitem{d'Enterria:2011kw}
D.~d'Enterria {\em et~al.}, ``{Constraints from the first LHC data on hadronic
  event generators for ultra-high energy cosmic-ray physics},''
  \href{http://dx.doi.org/10.1016/j.astropartphys.2011.05.002}{{\em
  Astropart.Phys.} {\bfseries 35} (2011) 98},
\href{http://arxiv.org/abs/1101.5596}{{\ttfamily arXiv:1101.5596 [astro-ph]}}.

\bibitem{Brun:1994aa}
R.~Brun, F.~Bruyant, F.~Carminati, S.~Giani, M.~Maire, A.~McPherson,
  G.~Patrick, and L.~Urban, ``{GEANT Detector Description and Simulation
  Tool},''
{\em Program Library Long Write-up W5013} (1994) .

\bibitem{2010arXiv1010.0632D}
G.~{D'Agostini}, ``{Improved iterative Bayesian unfolding},'' {\em ArXiv
  e-prints} (Oct., 2010) , \href{http://arxiv.org/abs/1010.0632}{{\ttfamily
  arXiv:1010.0632 [physics.data-an]}}.

\bibitem{Kaidalov:2009aw}
A.~B. Kaidalov and M.~G. Poghosyan, ``{Description of soft diffraction in the
  framework of reggeon calculus: Predictions for LHC},'' in {\em {Elastic and
  Diffractive Scattering. Proceedings, 13th International Conference, Blois
  Workshop, CERN, Geneva, Switzerland, June 29-July 3, 2009}}, pp.~91--98.
\newblock 2009.
\newblock \href{http://arxiv.org/abs/0909.5156}{{\ttfamily arXiv:0909.5156
  [hep-ph]}}.
\newblock
\url{https://inspirehep.net/record/832432/files/arXiv:0909.5156.pdf}.
\newblock

\bibitem{Abelev:2012sea}
{\bfseries ALICE} Collaboration, B.~Abelev {\em et~al.}, ``{Measurement of
  inelastic, single- and double-diffraction cross sections in proton--proton
  collisions at the LHC with ALICE},''
  \href{http://dx.doi.org/10.1140/epjc/s10052-013-2456-0}{{\em Eur. Phys. J.}
  {\bfseries C73} no.~6, (2013) 2456},
\href{http://arxiv.org/abs/1208.4968}{{\ttfamily arXiv:1208.4968 [hep-ex]}}.

\bibitem{Iancu:2003xm}
E.~Iancu and R.~Venugopalan,
  \href{http://dx.doi.org/10.1142/9789812795533_0005}{``{The Color glass
  condensate and high-energy scattering in QCD},''} in {\em In *Hwa, R.C. (ed.)
  et al.: Quark gluon plasma* 249-3363}.
\newblock 2003.
\newblock
\href{http://arxiv.org/abs/hep-ph/0303204}{{\ttfamily arXiv:hep-ph/0303204
  [hep-ph]}}.
\newblock

\bibitem{Koba:1972ng}
Z.~Koba, H.~B. Nielsen, and P.~Olesen, ``{Scaling of multiplicity distributions
  in high-energy hadron collisions},''
\href{http://dx.doi.org/10.1016/0550-3213(72)90551-2}{{\em Nucl.Phys.}
  {\bfseries B40} (1972) 317}.

\bibitem{Gelis:2009wh}
F.~Gelis, T.~Lappi, and L.~McLerran, ``{Glittering Glasmas},''
  \href{http://dx.doi.org/10.1016/j.nuclphysa.2009.07.004}{{\em Nucl.Phys.}
  {\bfseries A828} (2009) 149--160},
\href{http://arxiv.org/abs/0905.3234}{{\ttfamily arXiv:0905.3234 [hep-ph]}}.

\bibitem{McLerran:2008es}
L.~McLerran, \href{http://dx.doi.org/10.3204/DESY-PROC-2009-01/26}{``{A Brief
  Introduction to the Color Glass Condensate and the Glasma},''} in {\em
  {Proceedings, 38th International Symposium on Multiparticle Dynamics (ISMD
  2008): Hamburg, Germany, September 15-20, 2008}}, pp.~3--18.
\newblock 2009.
\newblock \href{http://arxiv.org/abs/0812.4989}{{\ttfamily arXiv:0812.4989
  [hep-ph]}}.
\newblock
\url{https://inspirehep.net/record/806434/files/arXiv:0812.4989.pdf}.
\newblock

\bibitem{Ghosh:2012xh}
P.~Ghosh, ``{Negative binomial multiplicity distribution in proton-proton
  collisions in limited pseudorapidity intervals at LHC up to $\sqrt{s} = 7$
  TeV and the clan model},''
  \href{http://dx.doi.org/10.1103/PhysRevD.85.054017}{{\em Phys.Rev.}
  {\bfseries D85} (2012) 054017},
\href{http://arxiv.org/abs/1202.4221}{{\ttfamily arXiv:1202.4221 [hep-ph]}}.

\bibitem{Giovannini:1998zb}
A.~Giovannini and R.~Ugoccioni, ``{Possible scenarios for soft and semihard
  components structure in central hadron hadron collisions in the TeV
  region},'' \href{http://dx.doi.org/10.1103/PhysRevD.59.094020,
  10.1103/PhysRevD.59.094020 10.1103/PhysRevD.69.059903,
  10.1103/PhysRevD.69.059903}{{\em Phys.Rev.} {\bfseries D59} (1999) 094020},
\href{http://arxiv.org/abs/hep-ph/9810446}{{\ttfamily arXiv:hep-ph/9810446
  [hep-ph]}}.

\end{thebibliography}\endgroup

\newpage
\appendix
\section{The ALICE collaboration}
\label{app:collab}



\begingroup
\small
\begin{flushleft}
S.~Acharya$^\textrm{\scriptsize 139}$,
D.~Adamov\'{a}$^\textrm{\scriptsize 96}$,
J.~Adolfsson$^\textrm{\scriptsize 34}$,
M.M.~Aggarwal$^\textrm{\scriptsize 101}$,
G.~Aglieri Rinella$^\textrm{\scriptsize 35}$,
M.~Agnello$^\textrm{\scriptsize 31}$,
N.~Agrawal$^\textrm{\scriptsize 48}$,
Z.~Ahammed$^\textrm{\scriptsize 139}$,
N.~Ahmad$^\textrm{\scriptsize 17}$,
S.U.~Ahn$^\textrm{\scriptsize 80}$
% ...
\renewcommand\labelenumi{\textsuperscript{\theenumi}~}

\section*{Affiliation notes}
\renewcommand\theenumi{\roman{enumi}}
\begin{Authlist}
\item \Adef{0}Deceased
\item \Adef{idp1828400}{Also at: Dipartimento DET del Politecnico di Torino, Turin, Italy}
\item \Adef{idp1847792}{Also at: Georgia State University, Atlanta, Georgia, United States}
\item \Adef{idp4144704}{Also at: M.V. Lomonosov Moscow State University, D.V. Skobeltsyn Institute of Nuclear, Physics, Moscow, Russia}
\item \Adef{idp4503248}{Also at: Department of Applied Physics, Aligarh Muslim University, Aligarh, India}
\item \Adef{idp5480656}{Also at: Institute of Theoretical Physics, University of Wroclaw, Poland}
\end{Authlist}

\section*{Collaboration Institutes}
\renewcommand\theenumi{\arabic{enumi}~}

$^{1}$A.I. Alikhanyan National Science Laboratory (Yerevan Physics Institute) Foundation, Yerevan, Armenia
\\
$^{2}$Benem\'{e}rita Universidad Aut\'{o}noma de Puebla, Puebla, Mexico
\\
$^{3}$Bogolyubov Institute for Theoretical Physics, Kiev, Ukraine
\\
$^{4}$Bose Institute, Department of Physics 
and Centre for Astroparticle Physics and Space Science (CAPSS), Kolkata, India
\\
$^{5}$Budker Institute for Nuclear Physics, Novosibirsk, Russia
\\
$^{6}$California Polytechnic State University, San Luis Obispo, California, United States
\\
$^{7}$Central China Normal University, Wuhan, China
\\
$^{8}$Centre de Calcul de l'IN2P3, Villeurbanne, Lyon, France
\\
$^{9}$Centro de Aplicaciones Tecnol\'{o}gicas y Desarrollo Nuclear (CEADEN), Havana, Cuba
\\
$^{10}$Centro de Investigaciones Energ\'{e}ticas Medioambientales y Tecnol\'{o}gicas (CIEMAT), Madrid, Spain
\\
$^{11}$Centro de Investigaci\'{o}n y de Estudios Avanzados (CINVESTAV), Mexico City and M\'{e}rida, Mexico
\\
$^{12}$Centro Fermi - Museo Storico della Fisica e Centro Studi e Ricerche ``Enrico Fermi', Rome, Italy
\\
$^{13}$Chicago State University, Chicago, Illinois, United States
\\
$^{14}$China Institute of Atomic Energy, Beijing, China
\\
$^{15}$COMSATS Institute of Information Technology (CIIT), Islamabad, Pakistan
\\
$^{16}$Departamento de F\'{\i}sica de Part\'{\i}culas and IGFAE, Universidad de Santiago de Compostela, Santiago de Compostela, Spain
\\
$^{17}$Department of Physics, Aligarh Muslim University, Aligarh, India
\\
$^{18}$Department of Physics, Ohio State University, Columbus, Ohio, United States
\\
$^{19}$Department of Physics, Pusan National University, Pusan, South Korea
\\
$^{20}$Department of Physics, Sejong University, Seoul, South Korea
\\
$^{21}$Department of Physics, University of Oslo, Oslo, Norway
\\
$^{22}$Department of Physics and Technology, University of Bergen, Bergen, Norway
\\
$^{23}$Dipartimento di Fisica dell'Universit\`{a} 'La Sapienza'
and Sezione INFN, Rome, Italy
\\
$^{24}$Dipartimento di Fisica dell'Universit\`{a}
and Sezione INFN, Cagliari, Italy
\\
$^{25}$Dipartimento di Fisica dell'Universit\`{a}
and Sezione INFN, Trieste, Italy
\\
$^{26}$Dipartimento di Fisica dell'Universit\`{a}
and Sezione INFN, Turin, Italy
\\
$^{27}$Dipartimento di Fisica e Astronomia dell'Universit\`{a}
and Sezione INFN, Bologna, Italy
\\
$^{28}$Dipartimento di Fisica e Astronomia dell'Universit\`{a}
and Sezione INFN, Catania, Italy
\\
$^{29}$Dipartimento di Fisica e Astronomia dell'Universit\`{a}
and Sezione INFN, Padova, Italy
\\
$^{30}$Dipartimento di Fisica `E.R.~Caianiello' dell'Universit\`{a}
and Gruppo Collegato INFN, Salerno, Italy
\\
$^{31}$Dipartimento DISAT del Politecnico and Sezione INFN, Turin, Italy
\\
$^{32}$Dipartimento di Scienze e Innovazione Tecnologica dell'Universit\`{a} del Piemonte Orientale and INFN Sezione di Torino, Alessandria, Italy
\\
$^{33}$Dipartimento Interateneo di Fisica `M.~Merlin'
and Sezione INFN, Bari, Italy
\\
$^{34}$Division of Experimental High Energy Physics, University of Lund, Lund, Sweden
\\
$^{35}$European Organization for Nuclear Research (CERN), Geneva, Switzerland
\\
$^{36}$Excellence Cluster Universe, Technische Universit\"{a}t M\"{u}nchen, Munich, Germany
\\
$^{37}$Faculty of Engineering, Bergen University College, Bergen, Norway
\\
$^{38}$Faculty of Mathematics, Physics and Informatics, Comenius University, Bratislava, Slovakia
\\
$^{39}$Faculty of Nuclear Sciences and Physical Engineering, Czech Technical University in Prague, Prague, Czech Republic
\\
$^{40}$Faculty of Science, P.J.~\v{S}af\'{a}rik University, Ko\v{s}ice, Slovakia
\\
$^{41}$Faculty of Technology, Buskerud and Vestfold University College, Tonsberg, Norway
\\
$^{42}$Frankfurt Institute for Advanced Studies, Johann Wolfgang Goethe-Universit\"{a}t Frankfurt, Frankfurt, Germany
\\
$^{43}$Gangneung-Wonju National University, Gangneung, South Korea
\\
$^{44}$Gauhati University, Department of Physics, Guwahati, India
\\
$^{45}$Helmholtz-Institut f\"{u}r Strahlen- und Kernphysik, Rheinische Friedrich-Wilhelms-Universit\"{a}t Bonn, Bonn, Germany
\\
$^{46}$Helsinki Institute of Physics (HIP), Helsinki, Finland
\\
$^{47}$Hiroshima University, Hiroshima, Japan
\\
$^{48}$Indian Institute of Technology Bombay (IIT), Mumbai, India
\\
$^{49}$Indian Institute of Technology Indore, Indore, India
\\
$^{50}$Indonesian Institute of Sciences, Jakarta, Indonesia
\\
$^{51}$INFN, Laboratori Nazionali di Frascati, Frascati, Italy
\\
$^{52}$INFN, Laboratori Nazionali di Legnaro, Legnaro, Italy
\\
$^{53}$INFN, Sezione di Bari, Bari, Italy
\\
$^{54}$INFN, Sezione di Bologna, Bologna, Italy
\\
$^{55}$INFN, Sezione di Cagliari, Cagliari, Italy
\\
$^{56}$INFN, Sezione di Catania, Catania, Italy
\\
$^{57}$INFN, Sezione di Padova, Padova, Italy
\\
$^{58}$INFN, Sezione di Roma, Rome, Italy
\\
$^{59}$INFN, Sezione di Torino, Turin, Italy
\\
$^{60}$INFN, Sezione di Trieste, Trieste, Italy
\\
$^{61}$Inha University, Incheon, South Korea
\\
$^{62}$Institut de Physique Nucl\'eaire d'Orsay (IPNO), Universit\'e Paris-Sud, CNRS-IN2P3, Orsay, France
\\
$^{63}$Institute for Nuclear Research, Academy of Sciences, Moscow, Russia
\\
$^{64}$Institute for Subatomic Physics of Utrecht University, Utrecht, Netherlands
\\
$^{65}$Institute for Theoretical and Experimental Physics, Moscow, Russia
\\
$^{66}$Institute of Experimental Physics, Slovak Academy of Sciences, Ko\v{s}ice, Slovakia
\\
$^{67}$Institute of Physics, Academy of Sciences of the Czech Republic, Prague, Czech Republic
\\
$^{68}$Institute of Physics, Bhubaneswar, India
\\
$^{69}$Institute of Space Science (ISS), Bucharest, Romania
\\
$^{70}$Institut f\"{u}r Informatik, Johann Wolfgang Goethe-Universit\"{a}t Frankfurt, Frankfurt, Germany
\\
$^{71}$Institut f\"{u}r Kernphysik, Johann Wolfgang Goethe-Universit\"{a}t Frankfurt, Frankfurt, Germany
\\
$^{72}$Institut f\"{u}r Kernphysik, Westf\"{a}lische Wilhelms-Universit\"{a}t M\"{u}nster, M\"{u}nster, Germany
\\
$^{73}$Instituto de Ciencias Nucleares, Universidad Nacional Aut\'{o}noma de M\'{e}xico, Mexico City, Mexico
\\
$^{74}$Instituto de F\'{i}sica, Universidade Federal do Rio Grande do Sul (UFRGS), Porto Alegre, Brazil
\\
$^{75}$Instituto de F\'{\i}sica, Universidad Nacional Aut\'{o}noma de M\'{e}xico, Mexico City, Mexico
\\
$^{76}$IRFU, CEA, Universit\'{e} Paris-Saclay, Saclay, France
\\
$^{77}$iThemba LABS, National Research Foundation, Somerset West, South Africa
\\
$^{78}$Joint Institute for Nuclear Research (JINR), Dubna, Russia
\\
$^{79}$Konkuk University, Seoul, South Korea
\\
$^{80}$Korea Institute of Science and Technology Information, Daejeon, South Korea
\\
$^{81}$KTO Karatay University, Konya, Turkey
\\
$^{82}$Laboratoire de Physique Corpusculaire (LPC), Clermont Universit\'{e}, Universit\'{e} Blaise Pascal, CNRS--IN2P3, Clermont-Ferrand, France
\\
$^{83}$Laboratoire de Physique Subatomique et de Cosmologie, Universit\'{e} Grenoble-Alpes, CNRS-IN2P3, Grenoble, France
\\
$^{84}$Lawrence Berkeley National Laboratory, Berkeley, California, United States
\\
$^{85}$Moscow Engineering Physics Institute, Moscow, Russia
\\
$^{86}$Nagasaki Institute of Applied Science, Nagasaki, Japan
\\
$^{87}$National and Kapodistrian University of Athens, Physics Department, Athens, Greece, Athens, Greece
\\
$^{88}$National Centre for Nuclear Studies, Warsaw, Poland
\\
$^{89}$National Institute for Physics and Nuclear Engineering, Bucharest, Romania
\\
$^{90}$National Institute of Science Education and Research, Bhubaneswar, India
\\
$^{91}$National Nuclear Research Center, Baku, Azerbaijan
\\
$^{92}$National Research Centre Kurchatov Institute, Moscow, Russia
\\
$^{93}$Niels Bohr Institute, University of Copenhagen, Copenhagen, Denmark
\\
$^{94}$Nikhef, Nationaal instituut voor subatomaire fysica, Amsterdam, Netherlands
\\
$^{95}$Nuclear Physics Group, STFC Daresbury Laboratory, Daresbury, United Kingdom
\\
$^{96}$Nuclear Physics Institute, Academy of Sciences of the Czech Republic, \v{R}e\v{z} u Prahy, Czech Republic
\\
$^{97}$Oak Ridge National Laboratory, Oak Ridge, Tennessee, United States
\\
$^{98}$Petersburg Nuclear Physics Institute, Gatchina, Russia
\\
$^{99}$Physics Department, Creighton University, Omaha, Nebraska, United States
\\
$^{100}$Physics department, Faculty of science, University of Zagreb, Zagreb, Croatia
\\
$^{101}$Physics Department, Panjab University, Chandigarh, India
\\
$^{102}$Physics Department, University of Cape Town, Cape Town, South Africa
\\
$^{103}$Physics Department, University of Jammu, Jammu, India
\\
$^{104}$Physics Department, University of Rajasthan, Jaipur, India
\\
$^{105}$Physikalisches Institut, Eberhard Karls Universit\"{a}t T\"{u}bingen, T\"{u}bingen, Germany
\\
$^{106}$Physikalisches Institut, Ruprecht-Karls-Universit\"{a}t Heidelberg, Heidelberg, Germany
\\
$^{107}$Physik Department, Technische Universit\"{a}t M\"{u}nchen, Munich, Germany
\\
$^{108}$Purdue University, West Lafayette, Indiana, United States
\\
$^{109}$Research Division and ExtreMe Matter Institute EMMI, GSI Helmholtzzentrum f\"ur Schwerionenforschung GmbH, Darmstadt, Germany
\\
$^{110}$Rudjer Bo\v{s}kovi\'{c} Institute, Zagreb, Croatia
\\
$^{111}$Russian Federal Nuclear Center (VNIIEF), Sarov, Russia
\\
$^{112}$Saha Institute of Nuclear Physics, Kolkata, India
\\
$^{113}$School of Physics and Astronomy, University of Birmingham, Birmingham, United Kingdom
\\
$^{114}$Secci\'{o}n F\'{\i}sica, Departamento de Ciencias, Pontificia Universidad Cat\'{o}lica del Per\'{u}, Lima, Peru
\\
$^{115}$SSC IHEP of NRC Kurchatov institute, Protvino, Russia
\\
$^{116}$Stefan Meyer Institut f\"{u}r Subatomare Physik (SMI), Vienna, Austria
\\
$^{117}$SUBATECH, IMT Atlantique, Universit\'{e} de Nantes, CNRS-IN2P3, Nantes, France
\\
$^{118}$Suranaree University of Technology, Nakhon Ratchasima, Thailand
\\
$^{119}$Technical University of Ko\v{s}ice, Ko\v{s}ice, Slovakia
\\
$^{120}$Technical University of Split FESB, Split, Croatia
\\
$^{121}$The Henryk Niewodniczanski Institute of Nuclear Physics, Polish Academy of Sciences, Cracow, Poland
\\
$^{122}$The University of Texas at Austin, Physics Department, Austin, Texas, United States
\\
$^{123}$Universidad Aut\'{o}noma de Sinaloa, Culiac\'{a}n, Mexico
\\
$^{124}$Universidade de S\~{a}o Paulo (USP), S\~{a}o Paulo, Brazil
\\
$^{125}$Universidade Estadual de Campinas (UNICAMP), Campinas, Brazil
\\
$^{126}$Universidade Federal do ABC, Santo Andre, Brazil
\\
$^{127}$University of Houston, Houston, Texas, United States
\\
$^{128}$University of Jyv\"{a}skyl\"{a}, Jyv\"{a}skyl\"{a}, Finland
\\
$^{129}$University of Liverpool, Liverpool, United Kingdom
\\
$^{130}$University of Tennessee, Knoxville, Tennessee, United States
\\
$^{131}$University of the Witwatersrand, Johannesburg, South Africa
\\
$^{132}$University of Tokyo, Tokyo, Japan
\\
$^{133}$University of Tsukuba, Tsukuba, Japan
\\
$^{134}$Universit\'{e} de Lyon, Universit\'{e} Lyon 1, CNRS/IN2P3, IPN-Lyon, Villeurbanne, Lyon, France
\\
$^{135}$Universit\'{e} de Strasbourg, CNRS, IPHC UMR 7178, F-67000 Strasbourg, France, Strasbourg, France
\\
$^{136}$Universit\`{a} degli Studi di Pavia, Pavia, Italy
\\
$^{137}$Universit\`{a} di Brescia, Brescia, Italy
\\
$^{138}$V.~Fock Institute for Physics, St. Petersburg State University, St. Petersburg, Russia
\\
$^{139}$Variable Energy Cyclotron Centre, Kolkata, India
\\
$^{140}$Warsaw University of Technology, Warsaw, Poland
\\
$^{141}$Wayne State University, Detroit, Michigan, United States
\\
$^{142}$Wigner Research Centre for Physics, Hungarian Academy of Sciences, Budapest, Hungary
\\
$^{143}$Yale University, New Haven, Connecticut, United States
\\
$^{144}$Yonsei University, Seoul, South Korea
\\
$^{145}$Zentrum f\"{u}r Technologietransfer und Telekommunikation (ZTT), Fachhochschule Worms, Worms, Germany
\endgroup

\end{document}